\documentclass[notitlepage,a4paper,aps,prd,onecolumn,superscriptaddress,nofootinbib,groupedaddress]{revtex4}

\usepackage{amsmath}
\usepackage{amsfonts,color}
\usepackage{amsmath}
\usepackage{amsthm}
\usepackage{pdfpages}
\usepackage{amsmath,cancel}
\usepackage{empheq}
\usepackage{amsfonts,color}
\usepackage{amssymb,float}
\usepackage{physics}
\usepackage{booktabs,multirow}
\usepackage{titlesec}
\usepackage{amsmath}

\titleformat{\paragraph}
{\normalfont\normalsize\bfseries}{\theparagraph}{1em}{}
\titlespacing*{\paragraph}
{0pt}{3.25ex plus 1ex minus .2ex}{1.5ex plus .2ex}

\usepackage{amsmath}
\usepackage[utf8]{inputenc}
\allowdisplaybreaks
\usepackage{color}
\usepackage{hyperref}
\hypersetup{
    colorlinks=true,
    linkcolor=blue,
    filecolor=magenta,      
   citecolor=blue
}

\usepackage{enumitem}

\usepackage{tikz}
\usetikzlibrary{shapes.geometric}
\usetikzlibrary{arrows.meta,arrows}

\begin{document}
\title{Stability of Poincaré gauge theory with cubic order invariants}

\author{Sebastian Bahamonde}
\email{sbahamondebeltran@gmail.com, sebastian.bahamonde@ipmu.jp}
\affiliation{Kavli Institute for the Physics and Mathematics of the Universe (WPI), The University of Tokyo Institutes
for Advanced Study (UTIAS), The University of Tokyo, Kashiwa, Chiba 277-8583, Japan.}

\author{Jorge Gigante Valcarcel}
\email{gigante.j.aa@m.titech.ac.jp}
\affiliation{Department of Physics, Tokyo Institute of Technology, 1-12-1 Ookayama, Meguro-ku, Tokyo 152-8551, Japan.}

\begin{abstract}

We analyse the stability of the vector and axial sectors of Poincaré gauge theory around general backgrounds in the presence of cubic order invariants defined from the curvature and torsion tensors, showing how the latter can in fact cancel out well-known instabilities arising from the quadratic curvature invariants of the theory and accordingly help in the construction of healthy models with both curvature and torsion. For this task, we introduce the most general parity preserving cubic Lagrangian with mixing terms of the curvature and torsion tensors, and find the relations of its coefficients to avoid a pathological behaviour from the vector and axial modes of torsion. As a result, on top of the gravitational constant of General Relativity and the mass parameters of torsion, our action contains 23 additional coupling constants controlling the dynamics of this field. As in the quadratic Poincaré gauge theory, we show that a further restriction on the cubic part of the action allows the existence of Reissner-Nordstr\"{o}m-like black hole solutions with dynamical torsion.

\end{abstract}

\maketitle

\section{Introduction}

The quest for a quantum theory of gravity beyond General Relativity (GR) remains one of the most long-standing challenges in theoretical physics. As is well-known, even though an effective field theory treatment for the metric tensor of GR provides a consistent way to compute probability amplitudes at low energies~\cite{Weinberg:1965nx,Jackiw:1968zza,Gross:1968in,Dunbar:1994bn,Donoghue:1996ma,Donoghue:1999qh,Cachazo:2014fwa,strominger2018lectures} (see~\cite{Wetterich:2017ixo,Wetterich:2018qsl} for nonperturvative regularisations in the infrared limit in the presence of a cosmological constant and effective scalar potentials),
its perturbative quantisation leads to ultraviolet divergences whose cancellation requires the appearance of infinite counter terms in the gravitational action~\cite{tHooft:1973bhk,tHooft:1974toh,Deser:1974zzd,Deser:1974cz,Deser:1974nb,Goroff:1985th}. By contrast, a generic extension of the Einstein-Hilbert action with quadratic curvature invariants renders gravity renormalisable~\cite{Stelle:1976gc}, but includes second order time derivatives of the metric tensor, which gives rise to a classical Hamiltonian unbounded from below~\cite{Ostrogradsky:1850fid,Buchbinder:1992rb,Woodard:2015zca}. The latter is in fact a feature naturally common in other extended theories of gravity~\cite{Damour:1992bt,Dvali:2000km,DeFelice:2009ak,Maartens:2010ar,Capozziello:2011et,Nojiri:2017ncd,Quiros:2019ktw,Lacombe:2023pmx}, representing a serious obstacle for their viability.

Different routes to try to tackle this issue have been proposed, both at the classical and at the quantum level~\cite{Lee:1969fy,Mannheim:2006rd,Bender:2007wu,Clunan:2009er,Zumalacarregui:2013pma,Salvio:2015gsi,Anselmi:2017ygm,Anselmi:2018kgz,Anselmi:2018tmf,Donoghue:2018izj,Donoghue:2019fcb,Donoghue:2021eto,Ganz:2020skf,Deffayet:2021nnt,Deffayet:2023wdg}. The most conservative procedure focuses on the cancellation of the instabilities by constraining the action to directly avoid the presence of any possible pathological term in the field equations, which for the case of metric theories of gravity with higher order curvature invariants strongly restricts the framework to Lanczos-Lovelock gravity~\cite{Lanczos:1938sf,Lovelock:1971yv,Padmanabhan:2013xyr}, hence to a theory that coincides with GR in four dimensions, up to a boundary term.

Thereby, it is clear that one of the most important aspects to analyse in the context of any extended theory of gravity concerns the stability of its propagating degrees of freedom. This task is especially relevant for theories that generally include a rich particle spectrum, as is the case of Metric-Affine Gravity (MAG), where the geometrical degrees of freedom correspond not only to the metric tensor but also to the torsion and nonmetricity tensors~\cite{Hehl:1994ue,Blagojevic:2013xpa,Cabral:2020fax}. In fact, besides the energy-momentum tensor of matter, this theory also presents a nontrivial hypermomentum tensor that describes additional properties of matter (namely the spin, dilation and shear currents of matter) and acts as a source of torsion and nonmetricity, thus constituting another extension of GR for all intents and purposes.

Nevertheless, one of the main challenges to unravel the physical properties and the phenomenological implications of the theory is the computational intractability due to the simultaneous presence of curvature, torsion and nonmetricity in any generic setup. This is especially why the majority of the stability analyses carried out so far in MAG have only considered models which are at most of quadratic order in the field
strength tensors of the theory~\cite{Neville:1978bk,Sezgin:1979zf,Sezgin:1981xs,Miyamoto:1983bf,Fukui:1984gn,Fukuma:1984cz,Battiti:1985mu,Kuhfuss:1986rb,Blagojevic:1986dm,Baikov:1992uh,Yo:1999ex,Yo:2001sy,Lin:2018awc,Jimenez:2019qjc,Percacci:2020ddy,Lin:2020phk,Marzo:2021iok,Baldazzi:2021kaf}. Remarkably, it has been recently shown that the cancellation of all the possible instabilities of the spin-1 modes from the quadratic action of MAG can only be fulfilled by reducing the parameter space of the quadratic curvature invariants from sixteen to five parameters~\cite{Jimenez-Cano:2022sds}, being the subcase of Weyl-Cartan geometry especially constrained to an Einstein-Proca theory for the Weyl vector of the nonmetricity tensor.

Following these lines, in this work we analyse the stability of the vector and axial modes of torsion around general backgrounds in the Poincaré gauge framework of MAG, showing that the well-known stability issues arising from the quadratic curvature invariants of the theory can be in fact suppressed in the presence of cubic order invariants defined from the curvature and torsion tensors. Indeed, as pointed out in~\cite{Cembranos:2017pcs,Bahamonde:2020fnq}, the relevance of the background space-time turns out to be essential to analyse the presence of instabilities in post-Riemannian geometry, which in any case can also be alleviated by the introduction in the gravitational action of higher order corrections depending on mixing terms. Thereby, here we follow this procedure and explicitly show how the introduction of cubic order corrections in the gravitational action can help in the construction of healthy theories with curvature and torsion.

We organise this paper as follows. First, in Sec.~\ref{sec:defs} we establish the definitions and conventions considered in our work for the main quantities described in the realm of Riemann-Cartan geometry, which constitutes a suitable framework to model a space-time endowed with curvature and torsion. Then, in Sec.~\ref{sec:quadraticPG} we briefly revisit the stability analysis performed around general backgrounds for the quadratic gravitational action of Poincaré gauge theory that simplifies to GR in the absence of torsion~\cite{Jimenez:2019qjc}, which clearly identifies the instabilities of the vector and axial sectors arising from the dynamics provided by the quadratic curvature invariants of the theory. From these results, we aim to alleviate the aforementioned pathologies by introducing cubic invariants constructed from the curvature and torsion tensors in the action. For this task, in Sec.~\ref{sec:cubictheory} we present the most general parity preserving cubic Lagrangian defined from these tensors, which in Sec.~\ref{sec:cubicstability} allows us to extend the stability analysis of the quadratic case and find an appropriate choice of the Lagrangian coefficients that cancels out the pathological terms in the vector and axial sectors of the theory. As an immediate application, it is a natural question to raise whether or not previous findings in the quadratic Poincaré gauge theory also hold in the presence of cubic order invariants. In particular, in Sec.~\ref{sec:solutions} we find that, similarly as in the quadratic case, a further restriction on the parameter space of the cubic action also allows the existence of Reissner-Nordstr\"{o}m-like black hole solutions with dynamical torsion. Finally, we end with summarising conclusions in Sec.~\ref{sec:conclusions}. For the sake of simplicity in the presentation, we relegate a list of cumbersome expressions and technical details to the Appendixes: in Appendix~\ref{sec:AppQuadratic} we present the contribution of the vector and axial modes of torsion to the quadratic curvature and torsion scalars, while in Appendix~\ref{sec:AppCubic} we show, for the cubic Lagrangian defined from the curvature and torsion tensors, the correspondence between the expressions derived from the split and unsplit forms of the torsion tensor.

We work in natural units $c=G=1$, and we consider the metric signature $(+,-,-,-)$. In addition, we use a tilde accent to denote quantities directly defined from the general affine connection, whereas their unaccented counterparts are defined from the Levi-Civita connection. Latin and Greek indices run from $0$ to $3$, referring to anholonomic and coordinate bases, respectively.

\section{Definitions and conventions}\label{sec:defs}

The formulation of gravity within a Riemann-Cartan manifold introduces the antisymmetric part of the affine connection as an additional property of the gravitational interaction. Specifically, the affine connection provides not only curvature changes in the geometry of the space-time but also a torsion deformation defined as
\begin{equation}
    T^{\lambda}\,_{\mu \nu}=2\tilde{\Gamma}^{\lambda}\,_{[\mu \nu]}\,.
\end{equation}

Then, the covariant derivative of an arbitrary vector $v^{\lambda}$ can be split into a Riemannian contribution and a contortion tensor
\begin{equation}
\tilde{\nabla}_{\mu}v^{\lambda}=\nabla_{\mu}v^{\lambda}+K^{\lambda}\,_{\rho\mu}v^{\rho}\,,
\end{equation}
with
\begin{equation}
    K^{\lambda}\,_{\rho\mu}=\frac{1}{2}\left(T^{\lambda}\,_{\rho\mu}-T_{\rho}\,^{\lambda}\,_{\mu}-T_{\mu}\,^{\lambda}\,_{\rho}\right)\,.
\end{equation}
Thereby, the torsion tensor modifies not only the expression of the covariant derivative but also its corresponding commutation relations, which provide the notion of the intrinsic curvature in terms of the affine connection
\begin{equation}
[\tilde{\nabla}_{\mu},\tilde{\nabla}_{\nu}]\,v^{\lambda}=\tilde{R}^{\lambda}\,_{\rho \mu \nu}\,v^{\rho}+T^{\rho}\,_{\mu \nu}\,\tilde{\nabla}_{\rho}v^{\lambda}\,,
\end{equation}
where
\begin{equation}\label{totalcurvature}
\tilde{R}^{\lambda}\,_{\rho \mu \nu}=\partial_{\mu}\tilde{\Gamma}^{\lambda}\,_{\rho \nu}-\partial_{\nu}\tilde{\Gamma}^{\lambda}\,_{\rho \mu}+\tilde{\Gamma}^{\lambda}\,_{\sigma \mu}\tilde{\Gamma}^{\sigma}\,_{\rho \nu}-\tilde{\Gamma}^{\lambda}\,_{\sigma \nu}\tilde{\Gamma}^{\sigma}\,_{\rho \mu}\,.
\end{equation}
In addition, in a Riemann-Cartan space-time the latter also provides one independent contraction for this tensor; namely, the Ricci tensor
\begin{equation}
\tilde{R}_{\mu\nu}=\tilde{R}^{\lambda}\,_{\mu \lambda \nu}\,,
\end{equation}
whose trace is identified with the scalar curvature
\begin{equation}
    \tilde{R}=g^{\mu\nu}\tilde{R}_{\mu\nu}\,,
\end{equation}
whereas the pseudotrace of the curvature tensor provides a linear pseudoscalar scalar quantity; namely, the Holst term
\begin{equation}
\ast\tilde{R}=\varepsilon^{\lambda\rho\mu\nu}\tilde{R}_{\lambda\rho\mu\nu}\,.
\end{equation}

For stability purposes and other phenomenological aspects, it is important to consider the irreducible decomposition of torsion as a tensor under the four-dimensional pseudo-orthogonal group~\cite{gambini1980einstein}:
\begin{equation}
T^{\lambda}\,_{\mu \nu}=\frac{1}{3}\left(\delta^{\lambda}\,_{\nu}T_{\mu}-\delta^{\lambda}\,_{\mu}T_{\nu}\right)+\frac{1}{6}\,\varepsilon^{\lambda}\,_{\rho\mu\nu}S^{\rho}+t^{\lambda}\,_{\mu \nu}\,,
\end{equation}
which presents vector, axial and tensor modes as:
\begin{align}\label{Tdec_1}
T_{\mu}&=T^{\nu}\,_{\mu\nu}\,,\\
S_{\mu}&=\varepsilon_{\mu\lambda\rho\nu}T^{\lambda\rho\nu}\,,\\
t_{\lambda\mu\nu}&=T_{\lambda\mu\nu}-\frac{2}{3}g_{\lambda[\nu}T_{\mu]}-\frac{1}{6}\,\varepsilon_{\lambda\rho\mu\nu}S^{\rho}\,.\label{Tdec3}
\end{align}

\section{Stability in quadratic Poincaré gauge theory}\label{sec:quadraticPG}

A gauge approach to gravity arises naturally when the unitary irreducible representations of relativistic particles labeled by their spin and mass are linked to the geometry of the space-time. Then, a gauge connection of the Poincar\'{e} group $ISO(1,3) = R^{4} \rtimes SO(1,3)$ can be introduced to describe the gravitational interaction as a gauge field of the external rotations and translations \cite{Hehl:1976kj,Obukhov:1987tz}. In particular, the curvature and torsion tensors acquire anholonomic expressions
\begin{eqnarray}
    F^{a}\,_{\mu\nu}&=&\partial_{\mu}e^{a}\,_{\nu}-\partial_{\nu}e^{a}\,_{\mu}+\omega^{a}\,_{b\mu}\,e^{b}\,_{\nu}-\omega^{a}\,_{b\nu}\,e^{b}\,_{\mu}\,,\\
    F^{a}\,_{b\mu\nu}&=&\partial_{\mu}\omega^{a}\,_{b\nu}-\partial_{\nu}\omega^{a}\,_{b\mu}+\omega^{a}\,_{c\mu}\,\omega^{c}\,_{b}\,_{\nu}-\omega^{a}\,_{c\nu}\,\omega^{c}\,_{b\mu}\,,
\end{eqnarray}
where the coframe $e^{a}\,_{\mu}$ and the gauge connection $\omega^{a}\,_{b\mu}$ satisfy the relations
\begin{eqnarray}
g_{\mu \nu}&=&e^{a}\,_{\mu}\,e^{b}\,_{\nu}\,\eta_{a b}\,,\\
\label{anholonomic_connection}
\omega^{a}\,_{b\mu}&=&e^{a}\,_{\lambda}\,e_{b}\,^{\rho}\,\tilde{\Gamma}^{\lambda}\,_{\rho \mu}+e^{a}\,_{\lambda}\,\partial_{\mu}\,e_{b}\,^{\lambda}\,,
\end{eqnarray}
being $\eta_{a b}$ the local Minkowski metric.

In order to introduce the dynamics of the gravitational field enhanced by torsion, the structure of the Poincaré group allows the definition of different nontrivial invariants constructed from curvature and torsion. Specifically, the most general parity preserving quadratic Lagrangian contains, besides the usual Ricci scalar, three torsion invariants and six additional curvature invariants, while the latter can be reduced to five in the gravitational action by the application of the Gauss-Bonnet theorem in four dimensions~\cite{Nieh:1979hf,Hayashi:1980bf}.

Following these lines, the gravitational action compatible with the gauge principles provides a vast number of kinetics and interactions for the irreducibles modes of torsion, which in general includes Ostrogradsky instabilities than can be directly avoided for certain combinations of the Lagrangian coefficients. In this sense, it has been shown that models containing only the Ricci scalar and the Holst term at quadratic order enables the propagation of healthy scalar and pseudoscalar modes for torsion, whereas the presence of the remaining quadratic curvature invariants spoils the stability of the spin-1 modes~\cite{Yo:1999ex,Yo:2001sy,Jimenez:2019qjc}. Therefore, here we focus on this pathological branch of the quadratic Poincaré gauge theory, which is given by an action that is reduced to GR in the absence of torsion:
\begin{align}
S=&\,\frac{1}{16\pi}\int
\Bigl[
-R-\frac{1}{2}\left(2c_{1}+c_{2}\right)\tilde{R}_{\lambda\rho\mu\nu}\tilde{R}^{\mu\nu\lambda\rho}+c_{1}\tilde{R}_{\lambda\rho\mu\nu}\tilde{R}^{\lambda\rho\mu\nu}+c_{2}\tilde{R}_{\lambda\rho\mu\nu}\tilde{R}^{\lambda\mu\rho\nu}+d_{1}\tilde{R}_{\mu\nu}\bigl(\tilde{R}^{\mu\nu}-\tilde{R}^{\nu\mu}\bigr)
\Bigr.\,
\nonumber\\
\Bigl.
&
+\frac{1}{2}m^2_TT_\mu T^\mu+\frac{1}{2}m^2_S S_\mu S^\mu +\frac{1}{2}m^2_tt_{\lambda\mu\nu}t^{\lambda\mu\nu}\Bigr]\sqrt{-g}\,d^4x\,,\label{Quadratic_Lag}
\end{align}
where $m_T$, $m_S$, $m_t$, $c_{1}$, $c_{2}$ and $d_{1}$ constitute the six Lagrangian coefficients of the quadratic order invariants.

For the analysis of the vector and axial sectors of the theory, we neglect then the tensor mode of torsion, which means that the curvature tensor acquires the form:
\begin{align}
		\tilde{R}_{\lambda\rho\mu\nu} =&\;R_{\lambda\rho\mu\nu}+\frac{2}{3}\left(g_{\lambda[\nu}\nabla_{\mu]}T_{\rho}-g_{\rho[\nu}\nabla_{\mu]}T_{\lambda}\right)+\frac{1}{6}\varepsilon_{\lambda\rho\sigma[\mu}\nabla_{\nu]}S^{\sigma}+\frac{2}{9}\left(2T_{[\lambda}g_{\rho][\nu}T_{\mu]}+g_{\lambda[\nu}g_{\mu]\rho}T_{\sigma}T^{\sigma}\right)\nonumber\\
        &+\frac{1}{72}\left(2S_{[\lambda}g_{\rho][\mu}S_{\nu]}+g_{\lambda[\mu}g_{\nu]\rho}S_{\sigma}S^{\sigma}\right)+\frac{1}{18}\left[2\varepsilon_{\mu\nu\sigma[\lambda}T_{\rho]}S^{\sigma}+\left(g_{\lambda[\nu}\varepsilon_{\mu]\rho\sigma\omega}-g_{\rho[\nu}\varepsilon_{\mu]\lambda\sigma\omega}\right)T^{\sigma}S^{\omega}\right]\,,\label{curvtensorTS}
\end{align}
whereas the corresponding Ricci tensor reads:
\begin{equation}
    \tilde{R}_{\mu\nu}=R_{\mu\nu}-\frac{1}{3}\left(2\nabla_{\nu}T_{\mu}+g_{\mu\nu}\nabla_{\lambda}T^{\lambda}\right)+\frac{1}{12}\varepsilon_{\lambda\rho\mu\nu}\nabla^{\lambda}S^{\rho}+\frac{2}{9}\left(T_{\mu}T_{\nu}-g_{\mu \nu}T_{\lambda}T^{\lambda}\right) +\frac{1}{72}\left(g_{\mu\nu}S_{\lambda}S^{\lambda}-S_{\mu}S_{\nu}\right)\,.\label{Ricci}
\end{equation}

From these quantities, it is straightforward to compute the corresponding quadratic invariants appearing in the gravitational action~\eqref{Quadratic_Lag}, which can be listed in Appendix~\ref{sec:AppQuadratic}. Then, by replacing their expressions in terms of the vector and axial modes, the Lagrangian reads
\begin{align}\label{preliminary_lag}
    16\pi\mathcal{L}=&\,-R+\frac{2}{9}\left(4 c_{1}+c_{2}+2 d_{1}\right) \nabla_{\mu}T_{\nu} \nabla^{\mu}T^{\nu}- \frac{2}{9}\left(4 c_{1}+c_{2}+2d_{1}\right)\nabla_{\mu}T_{\nu}\nabla^{\nu}T^{\mu}-\frac{1}{36}\left(4c_{1}+c_{2}+d_{1}\right) \nabla_{\mu}S_{\nu} \nabla^{\mu}S^{\nu}\nonumber\\
    &+\frac{1}{36}d_{1}\nabla_{\mu}S_{\nu}\nabla^{\nu}S^{\mu}+\frac{1}{72}\left(2c_{1}+5c_{2}\right)\nabla_{\mu}S^{\mu}\nabla_{\nu}S^{\nu}+\frac{1}{9}\left(4c_{1}+c_{2}+2d_{1}\right)\varepsilon^{\lambda\rho\mu\nu}\nabla_{\lambda}T_{\rho}\nabla_{\mu}S_{\nu}\nonumber\\
    &+\frac{1}{54}\left(4c_{1}+c_{2}\right)T^{\mu}S^{\nu}\nabla_{\mu}S_{\nu}+\frac{1}{54}\left(4c_{1}+c_{2}\right)T^{\mu}S^{\nu}\nabla_{\nu}S_{\mu}+\frac{1}{27}\left(2c_{2}-c_{1}\right)T_{\mu}S^{\mu}\nabla_{\nu}S^{\nu}\nonumber\\
    &-\frac{1}{162}\left(4c_{1}+ c_{2}\right)T_{\mu}T^{\mu}S_{\nu}S^{\nu}+\frac{1}{81}\left(2c_{2}-c_{1}\right)T_{\mu}S^{\mu}T_{\nu}S^{\nu}
    +\frac{1}{2}m^2_TT_\mu T^\mu+\frac{1}{2}m^2_S S_\mu S^\mu\,,
\end{align}
where the factor $\varepsilon^{\lambda\rho\mu\nu}\nabla_{\lambda}T_{\rho}\nabla_{\mu}S_{\nu}$ acts as a boundary term in the action and therefore it can be neglected from our calculations, since it does not contribute to the stability analysis. Furthermore, Expression~\eqref{preliminary_lag} can be rewritten by applying the following identities (up to boundary terms):
\begin{eqnarray}
    \int\nabla_\mu T^\mu \nabla_\nu T^\nu \sqrt{-g}\,d^4x&=&\int\left(\nabla_\mu T_\nu \nabla^\nu T^\mu+R_{\mu\nu}T^\mu T^\nu\right)\sqrt{-g}\,d^4x\,,\\
    \int\nabla_\mu S^\mu \nabla_\nu S^\nu \sqrt{-g}\,d^4x&=&\int\left(\nabla_\mu S_\nu \nabla^\nu S^\mu+R_{\mu\nu}S^\mu S^\nu\right)\sqrt{-g}\,d^4x\,,\\
    \int T_{\mu}S^{\mu}\nabla_{\nu}S^{\nu}\sqrt{-g}\,d^4x&=&-\int\left(T^{\mu}S^{\nu}\nabla_{\nu}S_{\mu}+S^{\mu}S^{\nu}\nabla_{\mu}T_{\nu}\right)\sqrt{-g}\,d^4x\,,\\
    \int T^{\mu}S^{\nu}\nabla_{\mu}S_{\nu}\sqrt{-g}\,d^4x&=&-\,\frac{1}{2}\int S_{\mu}S^{\mu}\nabla_{\nu}T^{\nu}\sqrt{-g}\,d^4x\,,
\end{eqnarray}
in order to conveniently display the different kinetic and interaction terms in the Lagrangian:
\begin{align}
    16\pi\mathcal{L}=&\,-R+\frac{1}{9}\left(4c_{1}+c_{2}+2d_{1}\right)F^{(T)}_{\mu\nu}F^{(T)\mu\nu}-\frac{1}{72}\left(4c_{1}+c_{2}+d_{1}\right)F^{(S)}_{\mu\nu}F^{(S)\mu\nu}+\frac{1}{24}\left(c_{2}-2c_{1}\right)\nabla_{\mu}S^{\mu}\nabla_{\nu}S^{\nu}\nonumber\\
    &+\frac{1}{27}\left(c_{1}-2c_{2}\right)S^{\mu}S^{\nu}\nabla_{\mu}T_{\nu}-\frac{1}{108}\left(4c_{1}+c_{2}\right)S_{\mu}S^{\mu}\nabla_{\nu}T^{\nu}+\frac{1}{18}\left(2c_{1}-c_{2}\right)T^{\mu}S^{\nu}\nabla_{\nu}S_{\mu}\nonumber\\
    &+\frac{1}{36}\left(4c_{1}+c_{2}\right)G_{\mu\nu}S^{\mu}S^{\nu}+\frac{1}{72}\left(4c_{1}+c_{2}\right)RS_{\mu}S^{\mu}-\frac{1}{162}\left(4c_{1}+ c_{2}\right)T_{\mu}T^{\mu}S_{\nu}S^{\nu}\nonumber\\
    &+\frac{1}{81}\left(2c_{2}-c_{1}\right)T_{\mu}S^{\mu}T_{\nu}S^{\nu}
     +\frac{1}{2}m^2_TT_\mu T^\mu+\frac{1}{2}m^2_S S_\mu S^\mu
    \,,\label{Lv}
\end{align}
where $G_{\mu\nu}=R_{\mu\nu}-(R/2)g_{\mu\nu}$ represents the Riemannian Einstein tensor, $F^{(T)}_{\mu\nu}=2\partial_{[\mu}T_{\nu]}$ and $F^{(S)}_{\mu\nu}=2\partial_{[\mu}S_{\nu]}$.

As pointed out in~\cite{Jimenez:2019qjc}, the pathological terms of the form $\left(\nabla S\right)^2$, $S^{2}\nabla T$, $TS\nabla S$ and $RS^{2}$ highly constrain the stability of the theory, since their cancellation in the gravitational action demands $c_{1}=c_{2}=0$, while the resulting kinetic terms of the vector and axial modes present opposite signs, which means that one of them unavoidably corresponds to a ghost field. Accordingly, the absence of these instabilities is achieved by switching off the dynamical character of the vector and axial modes, which reduces the theory to a model with nonpropagating torsion.

In the next sections, we shall see how these strong restrictions related to the stability of the vector and axial modes of torsion can be avoided by the addition of cubic order invariants constructed from the curvature and torsion tensors.

\section{Cubic Lagrangian}\label{sec:cubictheory}

Given the fact that curvature and torsion constitute rank-$4$ and rank-$3$ tensors, respectively, the most general parity preserving Lagrangian constructed from their cubic invariants presents two different branches: one provided by mixing terms involving these two tensors and another one by the curvature tensor alone. 

Specifically, the mixing terms related to the first branch depend linearly on the curvature tensor and quadratically on the torsion tensor. Therefore, its Lagrangian can be in general split into six different types, according to the possible combinations of the irreducible parts of torsion at quadratic order\footnote{A correspondence with the cubic Lagrangian expressed in terms of the unsplit torsion tensor can be found in Appendix~\ref{sec:AppCubic}.}:
\begin{equation}
    \mathcal{L}_{\rm curv-tors}^{(3)}=\mathcal{L}^{(3)}_{\tilde{R}TT}+\mathcal{L}^{(3)}_{\tilde{R}SS}+\mathcal{L}^{(3)}_{\tilde{R}tt}+\mathcal{L}^{(3)}_{\tilde{R}TS}+\mathcal{L}^{(3)}_{\tilde{R}Tt}+\mathcal{L}^{(3)}_{\tilde{R}St}\,,\label{cubicLagIrr}
\end{equation}
where
\begin{eqnarray}
    \mathcal{L}^{(3)}_{\tilde{R}TT}&=&h_{1}\tilde{R}_{\mu\nu}T^{\mu}T^{\nu}+h_{2}\tilde{R}T_{\mu}T^{\mu}\,,\\
    \mathcal{L}^{(3)}_{\tilde{R}SS}&=&h_{3}\tilde{R}_{\mu\nu}S^{\mu}S^{\nu}+h_{4}\tilde{R}S_{\mu}S^{\mu}\,,\\
    \mathcal{L}^{(3)}_{\tilde{R}tt}&=&h_5 \tilde{R}_{\lambda\rho\mu\nu}t_{\sigma}{}^{\lambda\rho}t^{\sigma\mu\nu}+h_6\tilde{R}_{\lambda\rho\mu\nu}t_{\sigma}{}^{\lambda\mu}t^{\sigma\rho\nu}+h_7\tilde{R}_{\lambda\rho\mu\nu}t^{\lambda\rho}{}_{\sigma}t^{\sigma\mu\nu}+h_{8}{} \tilde{R}_{\lambda\rho\mu\nu}t^{\lambda\mu}{}_{\sigma}t^{\sigma\rho\nu}+h_{9}{}\tilde{R}_{\lambda\rho\mu\nu}t^{\lambda\mu}{}_{\sigma}t^{\rho\nu\sigma}\nonumber\\
&&+\,h_{10}{}\tilde{R}_{\lambda\rho}t_{\mu\nu}{}^{\lambda}t^{\rho\mu\nu}+h_{11}{}\tilde{R}_{\lambda\rho}t_{\mu\nu}{}^{\lambda}t^{\mu\nu\rho}+h_{12}{}\tilde{R}t_{\lambda\rho\mu}t^{\lambda\rho\mu}\,,\\
    \mathcal{L}^{(3)}_{\tilde{R}TS}&=&h_{13}{}\varepsilon^{\lambda\rho\mu\nu}\tilde{R}_{\lambda\rho\mu\nu}T_{\sigma}S^{\sigma}+h_{14}{} \varepsilon_{\nu}{}^{\lambda\rho\sigma}\tilde{R}_{\lambda\rho\mu\sigma}T^{\mu}S^{\nu}+h_{15}{}\varepsilon^{\lambda\rho\mu\nu}\tilde{R}_{\lambda\rho}T_{\mu}S_{\nu}\,,\\
    \mathcal{L}^{(3)}_{\tilde{R}Tt}&=&h_{16}{}\tilde{R}_{\lambda\rho\mu\nu}T^{\nu}t^{\lambda\rho\mu}+h_{17}{} \tilde{R}_{\lambda\rho\mu\nu}T^{\rho}t^{\lambda\mu\nu}+h_{18}{}\tilde{R}_{\lambda\rho}T_{\mu}t^{\mu\lambda\rho}+h_{19}{} \tilde{R}_{\lambda\rho}T_{\mu}t^{\lambda\rho\mu}\,,\\
    \mathcal{L}^{(3)}_{\tilde{R}St}&=&h_{20}{}\varepsilon_{\alpha\rho\mu\nu}\tilde{R}_{\tau}{}^{\rho\mu\nu}S^{\gamma}t^{\alpha\tau}{}_{\gamma}+h_{21}{}\varepsilon_{\alpha\rho\mu\nu}\tilde{R}_{\tau}{}^{\rho\mu\nu}S^{\gamma}t_{\gamma}{}^{\alpha\tau}+h_{22}{}\varepsilon_{\alpha\rho}{}^{\mu\nu}\tilde{R}^{\rho}{}_{\mu\tau\nu}S^{\gamma}t_{\gamma}{}^{\alpha\tau}+h_{23}{}\varepsilon_{\alpha\rho}{}^{\mu\nu}\tilde{R}_{\gamma\mu\tau\nu}S^{\alpha}t^{\gamma\rho\tau}\nonumber\\
&&+\,h_{24}{}\varepsilon_{\alpha\rho}{}^{\mu\nu}\tilde{R}_{\gamma\mu\tau\nu} S^{\alpha}t^{\rho\tau\gamma}+h_{25}{}\varepsilon_{\alpha\rho\tau\mu}\tilde{R}^{\mu}{}_{\gamma}S^{\alpha}t^{\rho\tau\gamma}+h_{26}{}\varepsilon_{\lambda\rho\mu\nu} \tilde{R}^{\lambda\rho}S_{\sigma}t^{\sigma\mu\nu}\,.
\end{eqnarray}

On the other hand, the second branch displays all the independent cubic invariants constructed from the curvature tensor, including its respective traces:
\begin{equation}
    \mathcal{L}_{\rm curv}^{(3)}=\mathcal{L}^{(3)}_{\tilde{R}_{\lambda\rho\mu\nu}^3}\hspace{-3mm}+\mathcal{L}^{(3)}_{\tilde{R}_{\lambda\rho}^3}\hspace{-1mm}+\mathcal{L}^{(3)}_{\tilde{R}^3}+\mathcal{L}^{(3)}_{\tilde{R}_{\lambda\rho\mu\nu}^2\tilde{R}_{\alpha\beta}}\hspace{-2mm}+\mathcal{L}^{(3)}_{\tilde{R}_{\lambda\rho\mu\nu}\tilde{R}_{\alpha\beta}^2}\hspace{-2mm}+\mathcal{L}^{(3)}_{\tilde{R}_{\lambda\rho\mu\nu}^2\tilde{R}}+\mathcal{L}^{(3)}_{\tilde{R}_{\lambda\rho}^2\tilde{R}}\,,\label{cubicLagPureCurv}
\end{equation}
with
\begin{eqnarray}
\mathcal{L}^{(3)}_{\tilde{R}_{\lambda\rho\mu\nu}^3}&=&k_{1}^{} \tilde{R}_{\alpha \tau }{}^{\mu \nu } \tilde{R}^{\alpha \rho 
\tau \gamma } \tilde{R}_{\rho \gamma \mu \nu } + k_{2}^{} \tilde{R}_{
\alpha \tau }{}^{\mu \nu } \tilde{R}^{\alpha \rho \tau \gamma } \tilde{R}_{\rho \mu \gamma \nu } + k_{3}^{} \tilde{R}_{\alpha }{}^{\mu }{}_{\tau }{}^{\nu } \tilde{R}^{\alpha \rho \tau \gamma } \tilde{R}_{\rho \mu \gamma \nu } + k_{4}^{} \tilde{R}_{\alpha }{}^{\mu }{}_{\tau }{}^{\nu } \tilde{R}^{\alpha \rho \tau \gamma } \tilde{R}_{\rho \nu \gamma \mu }\nonumber\\
&&+\,k_{5}^{} \tilde{R}_{\alpha \rho }{}^{\mu \nu } \tilde{R}^{\alpha \rho \tau \gamma } \tilde{R}_{\tau \gamma \mu \nu } + k_{6}^{} \tilde{R}_{\alpha \rho }{}^{\mu \nu } \tilde{R}^{\alpha \rho \tau \gamma } \tilde{R}_{\tau \mu \gamma \nu } + k_{7}^{} \tilde{R}_{\alpha \tau }{}^{\mu \nu } \tilde{R}^{\alpha \rho \tau \gamma } \tilde{R}_{\gamma \mu \rho \nu } + k_{8}^{} \tilde{R}_{\alpha }{}^{\mu }{}_{\tau }{}^{\nu } \tilde{R}^{\alpha \rho \tau \gamma } \tilde{R}_{\gamma \nu \rho \mu } \nonumber\\
&&+\,k_{9}^{} \tilde{R}^{\alpha \rho \tau \gamma } \tilde{R}_{\tau }{}^{\mu }{}_{\alpha }{}^{\nu } \tilde{R}_{\gamma \nu \rho \mu } + k_{10}^{} \tilde{R}^{\alpha \rho \tau \gamma } \tilde{R}_{\tau \gamma }{}^{\mu \nu } \tilde{R}_{\mu \nu \alpha \rho } + k_{11}^{} \tilde{R}_{\alpha \tau }{}^{\mu \nu } \tilde{R}^{\alpha \rho \tau \gamma } \tilde{R}_{\mu \nu \rho \gamma }\,,\\
\mathcal{L}^{(3)}_{\tilde{R}_{\lambda\rho}^3}&=&k_{12}^{} \tilde{R}_{\rho }{}^{\mu } \tilde{R}^{\rho \tau } \tilde{R}_{\tau \mu } + k_{13}^{} \tilde{R}^{\rho \tau } \tilde{R}_{\tau }{}^{\mu } \tilde{R}_{\mu \rho }\,,\\
\mathcal{L}^{(3)}_{\tilde{R}^3}&=&k_{14}^{} \tilde{R}^3\,,\\
\mathcal{L}^{(3)}_{\tilde{R}_{\lambda\rho\mu\nu}^2\tilde{R}_{\alpha\beta}}&=&k_{15}^{} \tilde{R}^{\rho \tau } \tilde{R}_{\rho }{}^{\gamma \mu \nu 
} \tilde{R}_{\tau \gamma \mu \nu } + k_{16}^{} \tilde{R}^{\rho \tau } 
\tilde{R}_{\rho }{}^{\gamma \mu \nu } \tilde{R}_{\tau \mu \gamma \nu 
} + k_{17}^{} \tilde{R}^{\rho \tau } \tilde{R}_{\tau }{}^{\gamma \mu 
\nu } \tilde{R}_{\gamma \mu \rho \nu } + k_{18}^{} \tilde{R}^{\rho \tau } \tilde{R}_{\rho }{}^{\gamma \mu \nu } \tilde{R}_{\gamma \mu \tau 
\nu } \nonumber\\
&&+\,k_{19}^{} \tilde{R}^{\rho \tau } \tilde{R}_{\gamma \mu \tau \nu } \tilde{R}^{\gamma \mu }{}_{\rho }{}^{\nu } + k_{20}^{} \tilde{R}^{\rho \tau } \tilde{R}_{\gamma \nu \tau \mu } \tilde{R}^{\gamma \mu }{}_{\rho }{}^{\nu } + k_{21}^{} \tilde{R}^{\rho \tau } \tilde{R}_{\tau }{}^{\gamma \mu \nu } \tilde{R}_{\mu \nu \rho \gamma } + k_{22}^{} \tilde{R}^{\rho \tau } \tilde{R}_{\rho }{}^{\gamma \mu \nu } \tilde{R}_{\mu \nu \tau \gamma }\,,\\
\mathcal{L}^{(3)}_{\tilde{R}_{\lambda\rho\mu\nu}\tilde{R}_{\alpha\beta}^2}&=&k_{23}^{} \tilde{R}^{\rho \tau } \tilde{R}_{\gamma \mu } \tilde{R}_{\rho \tau }{}^{\gamma \mu } + k_{24}^{} \tilde{R}^{\rho \tau } \tilde{R}_{\gamma \mu } \tilde{R}_{\rho }{}^{\gamma }{}_{\tau }{}^{\mu } + k_{25}^{} \tilde{R}^{\rho \tau } \tilde{R}_{\mu \gamma } \tilde{R}_{\rho }{}^{\gamma }{}_{\tau }{}^{\mu } + k_{26}^{} \tilde{R}^{\rho \tau } \tilde{R}_{\mu \gamma } \tilde{R}_{\tau }{}^{\gamma }{}_{\rho }{}^{\mu }\,,\\
\mathcal{L}^{(3)}_{\tilde{R}_{\lambda\rho\mu\nu}^2\tilde{R}}&=&k_{27}^{} \tilde{R}_{\tau \gamma \mu \nu } \tilde{R}^{\tau \gamma \mu \nu } \tilde{R} + k_{28}^{} \tilde{R}_{\tau \mu \gamma \nu } \tilde{R}^{\tau \gamma \mu \nu } \tilde{R} + k_{29}^{} \tilde{R}^{\tau \gamma \mu \nu } \tilde{R}_{\mu \nu \tau \gamma } \tilde{R}\,,\\
\mathcal{L}^{(3)}_{\tilde{R}_{\lambda\rho}^2\tilde{R}}&=&k_{30}^{} \tilde{R}_{\gamma \mu } \tilde{R}^{\gamma\mu}\tilde{R}+ 
k_{31}^{} \tilde{R}^{\gamma \mu } \tilde{R}_{\mu \gamma } \tilde{R}\,.
\end{eqnarray}
Nevertheless, in four dimensions the third order Lanczos-Lovelock term
\begin{eqnarray}
    \mathcal{L}^{(3)}_{\rm LL}&=&\tilde{R}^{3}+2\tilde{R}^{\lambda\rho}{}_{\mu\nu}\tilde{R}^{\mu\nu}{}_{\sigma\omega}\tilde{R}^{\sigma\omega}{}_{\lambda\rho}+8\tilde{R}^{\lambda\rho}{}_{\mu\nu}\tilde{R}^{\mu\sigma}{}_{\rho\omega}\tilde{R}^{\nu\omega}{}_{\lambda\sigma}+16\tilde{R}^{\mu}{}_{\lambda}\tilde{R}^{\lambda}{}_{\rho}\tilde{R}^{\rho}{}_{\mu}\nonumber\\
    &&+\,24\tilde{R}^{\lambda\rho}{}_{\mu\nu}\tilde{R}^{\mu\nu}{}_{\rho\sigma}\tilde{R}^{\sigma}{}_{\lambda}+24\tilde{R}^{\lambda\rho}{}_{\mu\nu}\tilde{R}^{\mu}{}_{\lambda}\tilde{R}^{\nu}{}_{\rho}+3\tilde{R}\tilde{R}^{\lambda\rho}{}_{\mu\nu}\tilde{R}^{\mu\nu}{}_{\lambda\rho}-12\tilde{R}\tilde{R}^{\mu}{}_{\nu}\tilde{R}^{\nu}{}_{\mu}\,,
\end{eqnarray}
constitutes a topological invariant, which allows without any loss of generality one of the cubic curvature invariants to be eliminated from the four-dimensional action~\cite{Mardones:1990qc}.

In summary, in four dimensions both branches include 26 and 30 independent cubic invariants, respectively, which significantly enlarges the complexity of the theory. In any case, for the present stability analysis, the second branch is not relevant since it does not contain counter terms to remove the pathological interactions of the vector and axial sectors of the quadratic action. Hence, in the following section we will focus on the implications of the first branch on the stability problem.

\section{Stability of the vector and axial sectors at cubic order}\label{sec:cubicstability}

By including all the possible mixing terms of the curvature and torsion tensors of cubic order in the gravitational action, the dependence on the vector and axial modes of torsion acquires the following form:
\begin{align}
    16\pi\mathcal{L}=&\,-R+ \frac{1}{9}\left(4c_{1}+c_{2}+2d_{1}\right) F^{(T)}_{\mu\nu}F^{(T)\mu\nu}-\frac{1}{72} \left(4c_{1}+c_{2}+d_{1}\right) F^{(S)}_{\mu\nu}F^{(S)\mu\nu}+\frac{1}{24}\left(c_{2}-2c_{1}\right)\nabla_{\mu}S^{\mu}\nabla_{\nu}S^{\nu}\nonumber\\
    &+\frac{1}{54}\left[2c_{1}-4c_{2}-9\left(4h_{3}+6h_{13}+h_{14}\right)\right]S^{\mu}S^{\nu}\nabla_{\mu}T_{\nu}-\frac{1}{108}\left[4c_{1}+c_{2}+9\left(4h_{3}+24h_{4}+2h_{14}-h_{15}\right)\right] S_{\mu}S^{\mu}\nabla_{\nu}T^{\nu}\nonumber\\
    &+\frac{1}{18}\left[2c_{1}-c_{2}-3\left(6h_{13}+h_{14}-h_{15}\right)\right]T^{\mu}S^{\nu}\nabla_{\nu}S_{\mu}-2h_{2}T_{\mu}T^{\mu}\nabla_{\nu}T^{\nu}-\frac{2}{3}\left(h_{14}-h_{15}\right)\varepsilon^{\lambda\rho\mu\nu}T_{\lambda}S_{\rho}\partial_{\mu}T_{\nu}\nonumber\\
    &+\frac{1}{36}\left(4c_{1}+c_{2}+36h_{3}\right)G_{\mu\nu}S^{\mu}S^{\nu}+\frac{1}{72}\left[4c_{1}+c_{2}+36\left(h_{3}+2h_{4}\right)\right]RS_{\mu}S^{\mu}+h_{1}G_{\mu\nu}T^{\mu}T^{\nu}+\frac{1}{2}\left(h_{1}+2h_{2}\right)RT_{\mu}T^{\mu}\nonumber\\
    &-\frac{1}{648}\left[16 c_{1}+4c_{2}-9\left(h_{1}+3h_{2}-16h_{3}-48h_{4}-8h_{14}\right)\right]T_{\mu}T^{\mu}S_{\nu}S^{\nu}-\frac{2}{3}h_{2}T_{\mu} T^{\mu}T_{\nu}T^{\nu}+\frac{1}{24}h_{4}S_{\mu}S^{\mu}S_{\nu}S^{\nu}\nonumber\\
    &+\frac{1}{648}\left[16c_{2}-8c_{1}-9\left(h_{1}-16h_{3}-48h_{13}-8h_{14}\right)\right]T_{\mu}S^{\mu}T_{\nu}S^{\nu}
     +\frac{1}{2}m^2_TT_\mu T^\mu+\frac{1}{2}m^2_S S_\mu S^\mu
    \,.\label{vecaxsectorscubic}
\end{align}

As expected, the cubic Lagrangian introduces nonminimal couplings for these quantities. First, in contrast to the quadratic case, it explicitly provides interaction terms between the vector field and the metric tensor via the Riemannian Einstein tensor and the Ricci scalar, the latter exciting the longitudinal component of the vector mode with second time derivatives in the field equations and hence constituting a ghostly interaction~\cite{BeltranJimenez:2016rff}. Thus, we demand the following relation to suppress this issue in the vector sector around any general curved background:
\begin{equation}
    h_{2}=-\,\frac{h_{1}}{2}\,.\label{rel1}
\end{equation}
Conversely, self-interaction terms of the form $T^{2}\nabla T$ present a Galileon-like derivative coupling and do not generate Ostrogradsky instabilities~\cite{Jimenez:2019qjc,Delhom:2022vae}.

On the other hand, it turns out that the pathological terms of the form $\left(\nabla S\right)^2$, $S^{2}\nabla T$, $TS\nabla S$ and $RS^{2}$, driven by the axial sector, can be directly cancelled out from the action if the following relations are imposed on the Lagrangian coefficients:
\begin{equation}
    c_{2}=2c_{1}\,, \quad h_{3}= -\,\frac{1}{6}\left(c_{1}+6h_{13}\right)\,,\quad h_{4}= \frac{h_{13}}{2}\,,\quad h_{14}= -\,2h_{13}\,,\quad h_{15}=4h_{13}\,.\label{rel2}
\end{equation}
As can be seen, the pathologies affecting the axial sector in the quadratic Poincaré gauge theory arise then as the result of the truncation of the cubic invariants constructed from the curvature and torsion tensors in the action, which unavoidably demands $c_{1}=c_{2}=0$ and ultimately suppresses the dynamical character of torsion. However, by taking into account the cubic Lagrangian, it is possible to avoid this issue by imposing the relations~\eqref{rel2}. Furthermore, the ghostly instability associated with a wrong sign in the kinetic term of either the vector or the axial mode can also be circumvented if
\begin{equation}
    -\,\frac{d_1}{6} \leq c_{1} \leq -\,\frac{d_{1}}{3}\,, \quad d_1 \leq 0\,.\label{kincond}
\end{equation}
Note that the values $c_{1}=-\,d_{1}/3$ and $c_{1}=-\,d_{1}/6$ imply the absence of kinetic terms for the vector and axial modes, respectively, in which cases a combination of the form $T^{\mu}S^{\nu}F^{(S)}_{\mu\nu}$, arising from the previous terms of order $S^{2}\nabla T$ and $TS\nabla S$ by a suitable choice of the Lagrangian coefficients, would lead to a degenerate action and constitute an Ostrogradsky-free interaction. Nevertheless, in the present analysis we consider the general case with propagating vector and axial modes endowed with such kinetic terms and therefore we directly impose Expression~\eqref{rel2} to ensure stability.

It is worthwhile to stress that the term $\varepsilon^{\lambda\rho\mu\nu}T_{\lambda}S_{\rho}\partial_{\mu}T_{\nu}$ does not directly give rise to an Ostrogradsky instability. Indeed, by introducing Stückelberg fields, the corresponding scalar and pseudoscalar Stückelberg modes effectively appear as $T_\mu \rightarrow \partial_\mu T$ and $S_\mu \rightarrow \partial_\mu S$, which in the decoupling limit does not provide any pathological contribution from this term.

The stability of the potential can also be analysed by looking into the strong torsion regime, where the quartic contributions of order $T^4$, $S^4$ and $T^{2}S^{2}$ become dominant. Then, by taking into account the relations~\eqref{rel1} and~\eqref{rel2}, a straightforward computation of the eigenvalues associated with the matrix that describes the quartic order of the potential in the basis $\left(T_{\mu}T^{\mu}, S_{\mu}S^{\mu}, T_{\mu}S^{\mu}\right)$ yields
\begin{align}
    \lambda_{1}&=-\,\frac{1}{288}\left[48h_{1}+3h_{13}+\sqrt{5\left(461h_{1}^{2}+53h_{13}^{2}-64h_{1}h_{13}\right)}\,\right]\,,\label{eigen1}\\
    \lambda_{2}&=-\,\frac{1}{288}\left[48h_{1}+3h_{13}-\sqrt{5\left(461h_{1}^{2}+53h_{13}^{2}-64h_{1}h_{13}\right)}\,\right]\,,\label{eigen2}\\
    \lambda_{3}&=\frac{1}{72}\left(h_{1}-16h_{13}\right)\label{eigen3}\,.
\end{align}
The three eigenvalues are positive if
\begin{equation}
    h_{13} \leq \frac{h_{1}}{16} \leq \bigl(19+6\sqrt{10}\bigr)h_{13}\,, \quad h_{13} < 0\,,\label{quartic}
\end{equation}
which guarantees the stability in the strong torsion regime, while for vanishing values $h_{1}=h_{13}=0$ all the quartic contributions from the vector and axial modes directly disappear from the potential. This situation also differs from the case of the quadratic Poincaré gauge theory, where the quartic order of the potential for the action that is reduced to GR in the absence of torsion is completely driven by the coupling constants $c_{1}$ and $c_{2}$, preventing its stability or trivialisation unless $c_{1}=c_{2}=0$.

Following these lines, the stability conditions~\eqref{rel1}-\eqref{kincond} and~\eqref{quartic} ensure the absence in the vector and axial sectors of the well-known ghostly instabilities arising from the quadratic Poincaré gauge theory. Nevertheless, the complete gravitational action includes the corresponding kinetic and interaction terms of the tensor mode, whose stability analysis is yet to be fully understood~\cite{Jimenez:2019qjc,Jimenez-Cano:2022sds}. Overall, on top of the gravitational constant of Einstein's gravity and the mass parameters of torsion, the action displays $23$ coupling constants that are expected to be constrained in further stability analysis involving the tensor sector:
\begin{align}
    16\pi\mathcal{L}=&\,-R-2c_{1}\tilde{R}_{\lambda\rho\mu\nu}\tilde{R}^{\mu\nu\lambda\rho}+c_{1}\tilde{R}_{\lambda\rho\mu\nu}\tilde{R}^{\lambda\rho\mu\nu}+2c_{1}\tilde{R}_{\lambda\rho\mu\nu}\tilde{R}^{\lambda\mu\rho\nu}+d_{1}\tilde{R}_{\mu\nu}\bigl(\tilde{R}^{\mu\nu}-\tilde{R}^{\nu\mu}\bigr)+h_{1}\tilde{R}_{\mu\nu}T^{\mu}T^{\nu}
\Bigr.\,
\nonumber\\
\Bigl.
&-\frac{1}{2}h_{1}\tilde{R}T_{\mu}T^{\mu}-\frac{1}{6}\left(c_{1}+6h_{13}\right)\tilde{R}_{\mu\nu}S^{\mu}S^{\nu}+\frac{1}{2}h_{13}\tilde{R}S_{\mu}S^{\mu}+h_5 \tilde{R}_{\lambda\rho\mu\nu}t_{\sigma}{}^{\lambda\rho}t^{\sigma\mu\nu}+h_6\tilde{R}_{\lambda\rho\mu\nu}t_{\sigma}{}^{\lambda\mu}t^{\sigma\rho\nu}\nonumber\\
&+h_7\tilde{R}_{\lambda\rho\mu\nu}t^{\lambda\rho}{}_{\sigma}t^{\sigma\mu\nu}+h_{8}{} \tilde{R}_{\lambda\rho\mu\nu}t^{\lambda\mu}{}_{\sigma}t^{\sigma\rho\nu}+h_{9}{}\tilde{R}_{\lambda\rho\mu\nu}t^{\lambda\mu}{}_{\sigma}t^{\rho\nu\sigma}+h_{10}{}\tilde{R}_{\lambda\rho}t_{\mu\nu}{}^{\lambda}t^{\rho\mu\nu}+h_{11}{}\tilde{R}_{\lambda\rho}t_{\mu\nu}{}^{\lambda}t^{\mu\nu\rho}\nonumber\\
&+h_{12}{}\tilde{R}t_{\lambda\rho\mu}t^{\lambda\rho\mu}+h_{13}{}\varepsilon^{\lambda\rho\mu\nu}\tilde{R}_{\lambda\rho\mu\nu}T_{\sigma}S^{\sigma}-2h_{13}{}\varepsilon_{\nu}{}^{\lambda\rho\sigma}\tilde{R}_{\lambda\rho\mu\sigma}T^{\mu}S^{\nu}+4h_{13}{}\varepsilon^{\lambda\rho\mu\nu}\tilde{R}_{\lambda\rho}T_{\mu}S_{\nu}\nonumber\\
&+h_{16}{}\tilde{R}_{\lambda\rho\mu\nu}T^{\nu}t^{\lambda\rho\mu}+h_{17}{} \tilde{R}_{\lambda\rho\mu\nu}T^{\rho}t^{\lambda\mu\nu}+h_{18}{}\tilde{R}_{\lambda\rho}T_{\mu}t^{\mu\lambda\rho}+h_{19}{} \tilde{R}_{\lambda\rho}T_{\mu}t^{\lambda\rho\mu}+h_{20}{}\varepsilon_{\alpha\rho\mu\nu}\tilde{R}_{\tau}{}^{\rho\mu\nu}S^{\gamma}t^{\alpha\tau}{}_{\gamma}\nonumber\\
&+h_{21}{}\varepsilon_{\alpha\rho\mu\nu}\tilde{R}_{\tau}{}^{\rho\mu\nu}S^{\gamma}t_{\gamma}{}^{\alpha\tau}+h_{22}{}\varepsilon_{\alpha\rho}{}^{\mu\nu}\tilde{R}^{\rho}{}_{\mu\tau\nu}S^{\gamma}t_{\gamma}{}^{\alpha\tau}+h_{23}{}\varepsilon_{\alpha\rho}{}^{\mu\nu}\tilde{R}_{\gamma\mu\tau\nu}S^{\alpha}t^{\gamma\rho\tau}+h_{24}{}\varepsilon_{\alpha\rho}{}^{\mu\nu}\tilde{R}_{\gamma\mu\tau\nu} S^{\alpha}t^{\rho\tau\gamma}\nonumber\\
&+h_{25}{}\varepsilon_{\alpha\rho\tau\mu}\tilde{R}^{\mu}{}_{\gamma}S^{\alpha}t^{\rho\tau\gamma}+h_{26}{}\varepsilon_{\lambda\rho\mu\nu} \tilde{R}^{\lambda\rho}S_{\sigma}t^{\sigma\mu\nu}
     +\frac{1}{2}m^{2}_{T}T_{\mu} T^{\mu}+\frac{1}{2}m^{2}_{S}S_{\mu}S^{\mu}
+\frac{1}{2}m^{2}_{t}t_{\lambda\mu\nu}t^{\lambda\mu\nu}\,.\label{final_lag}
\end{align}

As a final remark, it is important to stress that in our conventions the absence of tachyonic instabilities in the three sectors demands the corresponding square masses of the irreducible modes of torsion to be nonnegative.

\section{Reissner-Nordstr\"{o}m-like black hole solutions}\label{sec:solutions}

As in the quadratic Poincaré gauge theory, the action given by the Lagrangian density~\eqref{final_lag} with cubic order invariants constructed from the curvature and torsion tensors admits Reissner-Nordstr\"{o}m-like black hole solutions with dynamical torsion. In particular, by considering a static and spherically symmetric setup~\cite{Hohmann:2019fvf}:
\begin{equation}\label{sph_metric}
    ds^{2}=\Psi_{1}(r)dt^{2} -\frac{dr^{2}}{\Psi_{2}(r)}-r^{2}d\vartheta^{2}-r^{2}\sin^{2}\vartheta \, d\varphi^{2}\,,
\end{equation}
\begin{align}\label{sph_torsion}
    T^t\,_{t r} &= t_{1}(r)\,, \quad T^r\,_{t r} = t_{2}(r)\,, \quad T^\vartheta\,_{t \vartheta} = T^\varphi\,_{t \varphi} = t_{3}(r)\,, \quad T^\vartheta\,_{r \vartheta} = T^\varphi\,_{r \varphi} =  t_{4}(r)\,, \\
    T^\vartheta\,_{t \varphi} &= T^\varphi\,_{\vartheta t} \sin^{2}\vartheta = t_{5}(r) \sin{\vartheta}\,, \quad T^\vartheta\,_{r \varphi} = T^\varphi\,_{\vartheta r} \sin^{2}\vartheta = t_{6}(r) \sin{\vartheta}\,, \\
    T^t\,_{\vartheta \varphi} &= t_{7}(r) \sin\vartheta\,, \quad T^r\,_{\vartheta \varphi} = t_{8}(r) \sin \vartheta\,,
\end{align}
as well as the stability conditions~\eqref{rel1} and~\eqref{rel2} for the vector and axial sectors, the solution of the Euler-Lagrange equations takes the form:
\begin{align}
    t_{1}(r)&=\frac{\Psi'(r)}{2\Psi(r)}+\frac{wr}{\Psi(r)}\,, \quad t_{2}(r)=t_{1}(r)\Psi(r)\,, \quad t_{3}(r)=-\,t_{4}(r)\Psi(r)\,, \quad t_{4}(r)=-\,\frac{1}{2r}-\frac{wr}{2\Psi(r)}\,,\label{tors1}\\
     t_{5}(r)&=-\,t_{6}(r)\Psi(r)\,,\quad  t_{6}(r)=\frac{N_{1}\kappa_{\rm s}}{r\Psi(r)}\,, \quad t_{7}(r)=\frac{N_{2}\kappa_{\rm s}r}{\Psi(r)}\,,\quad t_{8}(r)=t_{7}(r)\Psi(r)\,,\label{tors2}\\
     \Psi_{1}(r)&=\Psi_{2}(r)=\Psi(r)=1-\frac{2m}{r}
    +\left(2N_{1}-N_{2}\right) \left(N_{1}+N_{2}\right)\left[\frac{2N_{1}+N_{2}}{4N_{1}+N_{2}}d_{1}+2h_{25}\right]\frac{\kappa_{\rm s}^{2}}{3r^{2}}\,,\label{RNmetric}\\
m_{t}^{2}&=-\,\frac{6w\left(2N_1-N_2\right)\left[d_1 \left(2N_1+N_2\right)+\left(h_{25}-h_{6}\right)\left(4N_1+N_2\right)\right]}{\left(4N_1+N_2\right)\left(2N_1+5N_2\right)}\label{w_coeff}\,,
\end{align}
for the following set of Lagrangian coefficients:
\begin{eqnarray}\label{par1}
  c_1&=&\frac{c_2}{2}=-\,6h_3=-\,\frac{\left(2 N_1+N_2\right)}{2\left(4 N_1+N_2\right)}d_1 \,,\quad h_1= h_2=  h_4=  h_{13}=h_{14}=h_{15}=h_{16}=h_{17}=h_{18}=h_{19}=  0\,,\\
  h_{8}&=&\frac{1}{5}\left[9 d_1+10 h_{7}+9 h_{25}-4\left(5 h_{5}+h_{6}\right)+\frac{8 N_1 \left(4 d_1+9 h_{25}-9 h_{6}\right)}{2 N_1+5 N_2}-\frac{10N_{1}d_{1}}{4 N_1+N_2}\right]\,,\\
  h_{9}&=&-\,h_{10}=\frac{1}{10}\left[3 d_1+20 h_{7}+3 h_{25}-8\left(5 h_{5}+h_{6}\right)\right]+\frac{8 N_1 \left(4 d_1+9 h_{25}-9 h_{6}\right)}{5 \left(2 N_1+5 N_2\right)}+\frac{N_{1}d_{1}}{4 N_1+N_2} \,,\\
 h_{11}&=&2 h_{7}-4 h_{5}+\frac{9\left(2 N_1+N_2\right)\left[\left(2N_{1}+N_{2}\right)d_{1}+\left(4N_{1}+N_{2}\right)h_{25}\right]}{2\left(4 N_1+N_2\right)\left(2 N_1+5 N_2\right)}-\frac{\left(10 N_1+7 N_2\right)h_{6}}{2 N_1+5 N_2}\,,\\
  h_{12}&=&-\,\frac{1}{24} \left[3\left(d_{1}+4h_{7}+h_{25}-8 h_{5}-2h_{6}\right)+\frac{8 N_1 \left(4 d_1+9 h_{25}-9 h_6\right)}{2 N_1+5 N_2}+\frac{2N_{1}d_{1}}{4 N_1+N_2}\right]\,,\\
  h_{22}&=&\frac{1}{4} \left[4 h_{21}-2 d_1-8h_{20}-5 h_{25}-4 h_{26}+\frac{3 N_1 (2 d_1+3 h_{25})}{2N_1-N_2}+\frac{N_{1}h_{25}}{2 N_1+N_2}\right]\,,\\
  h_{23}&=&-\,2\left(d_1+2 h_{20}+h_{25}\right)+\frac{2 N_1 \left(2d_{1}+3h_{25}\right)}{2 N_1-N_2}+\frac{2N_{1}d_{1}}{4N_{1}+N_{2}}\,, \quad h_{24}=\frac{1}{2}\left(h_{23}+h_{25}\right)\,, \quad 
  m^2_T=m^2_S=0\,,\label{parF}
\end{eqnarray}
where $m$ and $\kappa_{\rm s}$ represent the parameters related to the mass and the spin charge of the solution, $w$ is related to the mass of torsion, and $N_{1}$ and $N_{2}$ are proportionality constants between the Lagrangian coefficients. As can be seen, the solution is analytical for general values of these constants, excluding the poles in the Expressions~\eqref{RNmetric}-\eqref{parF}; see Table~\ref{tab:special_cases} for the special cases that are not covered by such a generic solution. In any case, the additional stability condition~\eqref{kincond} sets the valid range of these constants for negative values of $d_{1}$ by the inequalities
\begin{equation}
    \frac{1}{3} \leq \frac{2N_{1}+N_{2}}{4N_{1}+N_{2}} \leq \frac{2}{3}\,,
\end{equation}
which allows the following possibilities:
\begin{itemize}
    \item $N_{2}=0$ and $-\,\infty < N_{1} < \infty$\,.
    \item $N_{2} > 0$ and $N_{1} \geq N_{2}/2$\,.
    \item $N_{2} > 0$ and $N_{1} \leq-\,N_{2}$\,.
    \item $N_{2} < 0$ and $N_{1} \geq-\,N_{2}$\,.
    \item $N_{2} < 0$ and $N_{1} \leq N_{2}/2$\,.
\end{itemize}
For $d_{1}=0$, the generic solution involves $c_{1}=c_{2}=0$, which trivialises the kinetic terms of torsion in the action. Only the special case $N_{2}=-\,4N_{1}$ provides a solution with nontrivial kinetic coefficients $c_{2}=2c_{1}=(1/3)\left(2h_{9}+3h_{25}-2h_{8}\right)$ (see Table~\ref{tab:special_cases}), but according to Expression~\eqref{kincond} the absence of an Ostrogradsky instability related to these terms in the resulting theory can be guaranteed if $c_{1}=c_{2}=0$ (i.e. if $h_{25}=(2/3)\left(h_{8}-h_{9}\right)$). Additionally, the special cases $N_{2}=-\,2N_{1}$ and $N_{2}=-\left(3\sqrt{2}+4\right)N_{1}$ also demand $d_{1}=0$ to satisfy this stability condition, whereas the cases $N_{2}=-\,N_{1}$, $N_{2}=2N_{1}$, $N_{2}=-\,2N_{1}/5$ and $N_{2}=\left(3\sqrt{2}-4\right)N_{1}$ are trivially allowed if $d_{1} \leq 0$.

Regarding the mass and potential terms, as in the solution of the quadratic Poincaré gauge theory~\cite{Cembranos:2017pcs}, the special case $N_{2}=2N_{1}$ includes a massive axial mode, whereas the cases $N_{2}=\bigl(\pm 3\sqrt{2}-4\bigr)N_{1}$ in Table~\ref{tab:special_cases} provide mass terms for both axial and tensor modes. Conversely, the generic solution~\eqref{tors1}-\eqref{w_coeff} and the rest of special cases are characterised by a massless axial mode and a massive tensor mode. All these configurations have a vanishing mass for the vector mode and must satisfy the tachyon-free condition by the avoidance of negative square masses for the other modes, which means $m^2_S \geq 0$ if $N_{2}=2N_{1}$, $m^2_S, m^2_t \geq 0$ if $N_{2}=\bigl(\pm 3\sqrt{2}-4\bigr)N_{1}$ and $m^2_t \geq 0$ in the rest of cases. On the other hand, only the special cases $N_{2}=\bigl(\pm 3\sqrt{2}-4\bigr)N_{1}$ appearing in Table~\ref{tab:special_cases} enable the quartic order of the potential introduced by the nontrivial coefficient $h_{13}$ in the action. Nevertheless, it is straightforward to check that the corresponding eigenvalues~\eqref{eigen1} and~\eqref{eigen2} present opposite signs in those cases, which means that the violation of the stability condition~\eqref{quartic} would unavoidably imply $h_{13}=0$, which as in all of the other cases trivialises the quartic order of the potential and additionally vanishes the mass of the axial mode. In this regard, it is important to note that even though such quartic order interactions generally arise from the quadratic and cubic invariants of the gravitational action~\eqref{vecaxsectorscubic}, they also naturally appear in the corresponding quartic Poincaré gauge theory as quartic torsion invariants with independent Lagrangian coefficients, in such a way that the presented solutions are not affected by their introduction in the action. Therefore, a trivial procedure to solve the aforementioned issue related to the stability of the quartic order of the potential consists of extending the gravitational action with quartic torsion invariants, fixing the respective Lagrangian coefficients to guarantee the correct sign in the eigenvalues, and to provide nontrivial masses for the axial and tensor modes in the special cases $N_{2}=\bigl(\pm 3\sqrt{2}-4\bigr)N_{1}$.

From the torsion components of the solution, it is straightforward to compute the corresponding vector, axial and tensor irreducible modes. In this sense, as in the quadratic Poincaré gauge model, we denote by a bar on top the quantities that introduce the dynamical effects of the spin charge in the geometry and by a circle on top the nondynamical part:
\begin{eqnarray}
T_{\mu}&=&\mathring{T}_{\mu}\,,\\
S_{\mu}&=&\bar{S}_{\mu}\,,\\
t_{\lambda\mu\nu}&=&\mathring{t}_{\lambda\mu\nu}+\bar{t}_{\lambda\mu\nu}\,,
\end{eqnarray}
where
\begin{align}
    \mathring{T}_{t}&=-\,\Psi(r)\mathring{T}_{r}=\frac{r\Psi'(r)+2\Psi(r)}{2r}+2rw\,,\quad \bar{S}_{t}=-\,\Psi(r)\bar{S}_{r}=\frac{2\left(N_{2}-2N_{1}\right)\kappa_{\rm s}}{r}\,,\\ 
     \mathring{t}^{r}{}_{t r}&=\Psi(r)\mathring{t}^{t}{}_{t r}=-\,2\mathring{t}^{\vartheta}{}_{t \vartheta}=-\,2\mathring{t}^{\varphi}{}_{t \varphi}=2\Psi(r)\mathring{t}^{\vartheta}{}_{r \vartheta}=2\Psi(r)\mathring{t}^{\varphi}{}_{r \varphi}=\frac{r\Psi'(r)-\Psi(r)}{3r}+\frac{wr}{3}\,,\\
    \bar{t}^{\,\vartheta}{}_{t\varphi}&=-\,\Psi(r)\bar{t}^{\,\vartheta}{}_{r\varphi}=-\,\bar{t}^{\,\varphi}{}_{t\vartheta}\sin^{2}\vartheta=\Psi(r)\bar{t}^{\,\varphi}{}_{r\vartheta}\sin^{2}\vartheta=-\,\frac{1}{2r^{2}}\,\bar{t}^{\,r}{}_{\vartheta\varphi}=-\,\frac{\Psi(r)}{2r^{2}}\,\bar{t}^{\,t}{}_{\vartheta\varphi}=-\,\frac{\left(N_{1}+N_{2}\right)\kappa_{\rm s}}{3 r}\,\sin\vartheta \,.
\end{align}
Then, the expressions of the antisymmetrised curvature tensor and of the antisymmetric Ricci tensor, which provide the kinetic terms of these parts, read
\begin{eqnarray}
 \bar{R}^{\lambda}\,_{[\mu \nu \rho]}&=&\frac{1}{6}\varepsilon^{\lambda}\,_{\sigma[\rho\nu}\nabla_{\mu]}\bar{S}^{\sigma}+\nabla_{[\mu}\bar{t}^{\lambda}\,_{\rho\nu]}-\frac{N_{1}}{9\left(2N_{1}-N_{2}\right)}\varepsilon_{\sigma\mu\nu\rho}\mathring{T}_{1}^{\lambda}\bar{S}^{\sigma}+\frac{2N_{1}+3N_{2}}{4\left(2N_{1}-N_{2}\right)}\varepsilon^{\lambda}\,_{\omega\sigma[\rho}\mathring{t}_{1}^{\sigma}\,_{\mu\nu]}\bar{S}^{\omega}\,,\\
     \bar{R}_{[\mu\nu]}&=&\frac{1}{12}\varepsilon^{\lambda}\,_{\sigma\mu\nu}\nabla_{\lambda}\bar{S}^{\sigma}+\frac{1}{2}\nabla_{\lambda}\bar{t}^{\lambda}\,_{\mu\nu}\,.
\end{eqnarray}

\begin{table}[H]
    \centering
  \bgroup
\def\arraystretch{3}  \scalebox{0.6}{\begin{tabular}{|c|c|l|}
    \hline
        Special case &  Metric function & \;\;\;\;\;\;\;\;\;\;\;\;\;\;\;\;\;\;\;\;\;\;\;\;\;\;\;\;\;\;\;\;\;\;\;\;\;\;\;\;\;\;\;\;\;\;\;\;\;\;\;\;\;\;\;\;\;\;\;\;\;\;\;\;\;\;\;\;\;\;\;\;\;\;\;\;\;\;\;\;\;Lagrangian coefficients \\
        \hline
        \multirow{3}{*}{\vspace{-3mm}$N_{2}=-\,4N_{1}$}  & \multirow{3}{*}{\vspace{-3mm}$\displaystyle \Psi(r)=1-\frac{2m}{r}+\frac{2\left[2\left(h_{9}-h_{8}\right)-3h_{25}\right]N_{1}^{2}\kappa_{\rm s}^{2}}{r^2} $} & $\displaystyle c_{1}=\frac{1}{2}\left[4h_{7}+h_{25}-2\left(4h_{5}+h_{6}+h_{11}\right)\right],d_1= h_{1}= h_{13}= h_{16}= h_{17}= h_{18}=h_{19}= 0\,, h_{8}=2\left(2h_{5}+h_{6}+h_{11}-h_{7}\right),$\\ 
       & &$ \displaystyle h_{9}=-\,h_{10}=4h_{7}-8h_{5}-h_{6}-h_{11}\,,h_{12}=\frac{1}{12}\left(16h_{5}+2h_{6}+h_{11}-8h_{7}\right), h_{22}=h_{21}-2h_{20}-h_{25}-h_{26}\,,$   \\
       & & $ \displaystyle h_{23}=2\left(2h_{7}-4h_{5}-h_{6}-h_{11}-2h_{20}\right),h_{24}=\frac{1}{2}\left(h_{23}+h_{25}\right),$ 
       $\displaystyle m_S=m_T=0, m_t^2=-2 w  (8 h_{5}+h_{6}-4 h_{7}+2 h_{9})$
       \\[2ex]\hline
        \multirow{4}{*}{\vspace{8mm}$N_{2}=-\,2N_{1}/5$\,}  & \multirow{4}{*}{\vspace{8mm}$\displaystyle \Psi(r)=1-\frac{2m}{r}+\frac{8\left(2d_{1}+9h_{25}\right) N_{1}^{2}\kappa_{\rm s}^{2}}{75 r^{2}}$} & $\displaystyle c_1= -\,\frac{2}{9} d_1\,,h_{1}=h_{13}=h_{16}= h_{17}=h_{18}= h_{19}=0\,,h_{6}= \frac{4 d_1}{9}+h_{25}\,, h_{8}=\frac{8d_{1}}{9}+4h_{5}+2h_{11}+2h_{25}-2h_{7}\,,$\\ 
       & &$ \displaystyle h_{9}=-\,h_{10}=\frac{2d_{1}}{9}+4h_{5}+2h_{11}+\frac{h_{25}}{2}-2h_{7}\,, h_{12}=-\,\frac{d_{1}}{27}+\frac{1}{12}\left(4h_{7}-8h_{5}-5h_{11}-h_{25}\right)\,,h_{23}=\frac{2d_{1}}{9}+\frac{h_{25}}{2}-4h_{20}\,,$   \\
       & & $\displaystyle  h_{22}=\frac{d_{1}}{8}+h_{21}-2h_{20}-\frac{5h_{25}}{32}-h_{26}\,, h_{24}=\frac{d_{1}}{9}+\frac{1}{4}\left(3h_{25}-8h_{20}\right),$
       $\displaystyle m_S=m_T=0,m_t^2=-\frac{1}{18} w  (4 d_1+9 (h_{25}+8 h_{5}-4 h_{7}+2 h_{9}))$
       \\[2ex]\hline
       \multirow{3}{*}{\vspace{7mm}$N_{2}=2N_{1}$}  & \multirow{3}{*}{\vspace{7mm}$\displaystyle \Psi(r)=1-\frac{2 m}{r}$} & $ \displaystyle c_1= -\,\frac{d_1}{3}\,,h_{1}=h_{13}=h_{16}=h_{17}=h_{18}=h_{19}=0\,,h_{9}=-\,h_{10}=h_{11}=-\,4h_{12}=h_{8}\,, h_{8}=2\left(h_{7}-h_{6}-2h_{5}\right),$\\ 
       & &$ \displaystyle h_{22}=h_{21}+\frac{1}{8}\left(3h_{23}-4h_{20}-5h_{25}-8h_{26}\right),h_{24}=\frac{1}{2}\left(h_{23}+h_{25}\right),$ 
       $\displaystyle m_T=m_t=0$
       \\[2ex]\hline
       \multirow{3}{*}{\vspace{7mm}$N_{2}=-\,2N_{1}$}  & \multirow{3}{*}{\vspace{7mm}$\displaystyle \Psi(r)=1-\frac{2m}{r}$} & $ \displaystyle c_{1}= h_{1}= h_{13}=h_{16}=h_{17}=h_{19}=h_{25}= 0\,,h_{8}=h_{9}=-\,h_{10}=h_{6}+2h_{7}-4h_{5}\,, h_{11}=\frac{1}{2}\left(4h_{7}-8h_{5}-h_{6}\right),$\\ 
       & &$ \displaystyle h_{12}=h_{5}-\frac{1}{8}\left(h_{6}+4h_{7}\right),h_{23}=-\,4h_{20}\,,h_{24}= \frac{h_{23}}{2}\,,$
       $\displaystyle m_S=m_T=0, m_t^2=-3w\,h_6$
       \\[2ex]\hline
       \multirow{2}{*}{\vspace{-3mm}$N_{2}=-\,N_{1},$}  & \multirow{3}{*}{$\displaystyle \Psi(r)=1-\frac{2 m}{r}$} & $ \displaystyle c_1= -\,\frac{d_1}{6}\,,h_{1}=h_{13}=h_{19}=0\,,h_{9}=\frac{1}{2}\left(4h_{5}+3h_{8}-2h_{7}-4h_{6}\right), h_{10}=\frac{1}{3}\left(h_{6}+2h_{7}+12h_{12}-4h_{5}-h_{8}\right),$\\ 
      \multirow{2}{*}{\vspace{-3mm}$h_{17}\neq0$}   & &$ \displaystyle h_{11}=\frac{1}{6}\left(14h_{6}+10h_{7}-20h_{5}-11h_{8}-48h_{12}\right), h_{16}=2h_{17}=-\,h_{18}\,, h_{22}=\frac{1}{12}\left(d_{1}+12h_{21}-3h_{6}-24h_{20}-12h_{26}\right),$\\
       & &$ \displaystyle h_{23}=-\,4h_{20}\,, h_{24}=\frac{1}{6}\left(3h_{6}-d_{1}-12h_{20}\right), h_{25}=\frac{1}{3}\left(3h_{6}-d_{1}\right),$
       $\displaystyle m_S=m_T=0,  m_t^2=-\frac{1}{3} w  (27 h_{10}+16 h_{5}+2 h_{6}-8 h_{7}+31 h_{9})$
       \\[2ex]\hline
      \multirow{4}{*}{\vspace{-18mm}$N_{2}=\bigl(\pm 3\sqrt{2}-4\bigr)N_{1},$}  & \multirow{4}{*}{\vspace{-8mm}$\displaystyle \Psi(r)=1-\frac{2m}{r}+\frac{1}{r^2}\bigr[d_{1}\bigl(\pm 13 \sqrt{2}-18\bigr)$}&
      $\displaystyle c_1=-\,\frac{1}{6}\bigl(3\mp\sqrt{2}\bigr)d_{1}\,,h_{1}=h_{19}=0\,,h_{6}=\frac{1}{3} \bigl(\pm 2 \sqrt{2}-1\bigr) h_{11}+\bigl(\pm 108 \sqrt{2}+87\bigr)h_{13}\pm \frac{1}{\sqrt{2}}\bigl(4h_{20}-h_{25}\bigr) \pm \sqrt{2} h_{24}+\frac{1}{3}\bigl(1\mp 2 \sqrt{2}\bigr)h_{9}\,,$\\
    \multirow{4}{*}{\vspace{-20mm}$h_{17}\neq0$}  &  \multirow{4}{*}{\vspace{-7mm}$\displaystyle +\,3h_{17} \bigl(\pm 60\sqrt{2}-85\bigr)$}&$\displaystyle h_8=h_{9}-27\bigl(\pm\sqrt{2}+2\bigr) h_{13}\pm 3\sqrt{2}h_{20}\pm\frac{3h_{23}}{2\sqrt{2}}\,,$\\
    && $\displaystyle h_{9}=\frac{1}{248}\bigl[4\bigl(\pm 13\sqrt{2}-11\bigr)d_{1}+32\bigl(\pm 3\sqrt{2}+7\bigr)h_{11}+12\bigl(\pm 4\sqrt{2}-1\bigr)\bigl(8h_{5}+h_{25}-4h_{7}\bigr)+3\bigl(\pm 659\sqrt{2}-1180\bigr)h_{17}\bigr],$\\
      && $\displaystyle h_{10}=\frac{3}{8} \bigl(\pm 2 \sqrt{2}-3\bigr)h_{17}-h_{9}\,, h_{12}=\frac{1}{84}\bigl(\pm 4 \sqrt{2}+23\bigr)h_{10}+\frac{3}{28}\bigl(\pm 118 \sqrt{2}+199\bigr)h_{13}+\frac{1}{84}\bigl(\pm 2\sqrt{2}+1\bigr) \bigl(2h_{8}-3h_{6}\bigr),$\\ 
      &\multirow{2}{*}{\vspace{17mm}$\displaystyle \,\,\,\,\,\,\,\,\,\,\,+\,6h_{25} \bigl(\pm 3 \sqrt{2}-4\bigr)\bigr]N_{1}^{2}\kappa_{\rm s}^{2}$}& $ \displaystyle h_{13}=\frac{1}{12}\bigl(\pm 2\sqrt{2}-3\bigr)h_{17}\,, h_{16}=-\,24 \bigl(\pm 2\sqrt{2}+3\bigr)h_{13}\,, h_{18}=-\,2\bigl(\pm 3\sqrt{2}-5\bigr)h_{17}\,,$\\ 
       && $\displaystyle h_{22}= \frac{1}{56}\bigl[\pm 14 \sqrt{2} d_1+\bigl(249\mp 99\sqrt{2}\bigr)h_{17}-56\bigl(2h_{20}-h_{21}+h_{26}\bigr)+\bigl(\pm 24\sqrt{2}-26\bigr)h_{25}\bigr],\displaystyle h_{23}=\pm 6\sqrt{2}h_{13}+2h_{24}-h_{25}\,,$\\
       && $\displaystyle h_{24}=\frac{1}{12} \bigl\{2\bigl(\pm 3\sqrt{2}-2\bigr)d_{1}+9\bigl(13\mp 7\sqrt{2}\bigr)h_{17}+6\bigl[\bigl(1\pm\sqrt{2}\bigr)h_{25}-4h_{20}\bigr]\bigr\},\displaystyle m_{T}=0\,, m_{S}^{2}=6h_{13}w\,, m_{t}^{2}=\frac{3}{2}\bigl(4h_{10}-16h_{12}-27h_{13}\bigr)w$ 
      \\[2ex]\hline
    \end{tabular}}\egroup
    \caption{Solutions for special cases\protect\footnotemark.}
    \label{tab:special_cases}
\end{table}

\footnotetext{For simplicity in the presentation, we do not include the values of the Lagrangian coefficients of the solutions set by the stability conditions nor their torsion components, which are directly provided for each special case by the Expressions~\eqref{rel1},~\eqref{rel2},~\eqref{tors1} and~\eqref{tors2}.}

\section{Conclusions}\label{sec:conclusions}

The search of ghost and tachyon free gravitational models with dynamical torsion in the framework of quadratic Poincaré gauge theory is in general an arduous task~\cite{Neville:1978bk,Sezgin:1979zf,Sezgin:1981xs,Miyamoto:1983bf,Fukui:1984gn,Fukuma:1984cz,Battiti:1985mu,Kuhfuss:1986rb,Blagojevic:1986dm,Baikov:1992uh,Yo:1999ex,Yo:2001sy,Lin:2018awc,Jimenez:2019qjc}. In fact, an examination on the inherent nonlinearity of the interactions involving the vector and axial modes of torsion is enough to reveal a significant difficulty in finding a healthy model within this framework, beyond the biscalar subclass that switches off the propagation of the pure vector, axial and tensor modes of torsion~\cite{Yo:1999ex,Yo:2001sy,Jimenez:2019qjc}. Although the latter offers a safe context and an interesting phenomenology to study the physical implications of the scalar and pseudoscalar degrees of freedom of torsion~\cite{Yo:2006qs,Shie:2008ms,Chen:2009at,Ao:2011kc,tseng2012scalar,Geng:2013hp,geng2014scalar,Lu:2016bcx,Zhang:2019mhd,Zhang:2019xek,delaCruzDombriz:2021nrg,Casado-Turrion:2023omz}, it is reasonable to investigate different routes to overcome this shortcoming concerning the rest of sectors of the theory.

Thereby, in this work we have explicitly shown how the introduction of cubic order invariants defined from the curvature and torsion tensors in the action allows the cancellation of the gravitational instabilities present in the vector and axial sectors around any general curved background. The parameter space of the resulting Lagrangian includes, on top of the gravitational constant of GR and the mass parameters of torsion, $23$ coupling constants that control the dynamics of this field. In this sense, the pathologies affecting the quadratic Poincaré gauge theory can then be understood as the result of the truncation of the cubic invariants constructed from the curvature and torsion tensors in the action, whose further implications and restrictions are yet to be clarified by analysing the stability of the tensor sector (including its kinetics and interactions with the vector and axial modes of torsion) in a general background and other consistency requirements, such as the existence of solutions with a suitable asymptotic behavior~\cite{Chen:1988mz}. Indeed, we have also pointed out the existence of Reissner-Nordstr\"{o}m-like black hole configurations, in line with the quadratic Poincaré gauge theory, which constitute exact solutions of such a theory with stable vector and axial sectors under different restrictions of the Lagrangian coefficients.

As natural generalisations of these results, it is important to extend the present stability analysis to general metric-affine geometries, in order to perform a thorough examination on the stability of both torsion and nonmetricity fields~\cite{BeltranJimenez:2019acz,Percacci:2020ddy,BeltranJimenez:2020sqf,Lin:2020phk,Marzo:2021iok,Baldazzi:2021kaf,Jimenez-Cano:2022sds}. From a mathematical point of view, such an extension is highly nontrivial since it provides a much larger number of independent cubic invariants constructed from the curvature, torsion and nonmetricity tensors in the gravitational action. For this reason, this result will be presented separately in a future work. Likewise, as in the quadratic Poincaré gauge theory, it is possible to include parity violating interactions, which complement the even parity gravitational action and might be relevant at astrophysical and cosmological scales~\cite{Baekler:2011jt,Karananas:2014pxa,Obukhov:2020hlp}. We consider such an extension also deserves further investigation in future works.

\bigskip
\bigskip
\noindent
\section*{Acknowledgements}

We would like to thank Alejandro Jiménez Cano and Francisco José Maldonado Torralba for helpful discussions. S.B. is supported by “Agencia Nacional de Investigación y Desarrollo” (ANID), Grant “Becas Chile postdoctorado
al extranjero” No. 74220006. J.G.V. is supported by JSPS Postdoctoral Fellowships for Research in Japan and KAKENHI Grant-in-Aid for Scientific Research No. JP22F22044.

\newpage

\appendix

\section{Quadratic scalars with vector and axial contributions}\label{sec:AppQuadratic}

From the Expressions~\eqref{curvtensorTS} and~\eqref{Ricci} containing the corrections provided by the vector and axial modes of torsion to the curvature and Ricci tensors, it is possible to obtain six independent quadratic curvature scalars:
\begin{align}
    \tilde{R}_{\lambda\rho\mu\nu}\tilde{R}^{\lambda\rho\mu\nu}=&\;R_{\lambda\rho\mu\nu}R^{\lambda\rho\mu\nu}+\frac{8}{9}\nabla_{\mu}T_{\nu}\nabla^{\mu}T^{\nu}+\frac{4}{9}\nabla_{\mu}T^{\mu} \nabla_{\nu}T^{\nu}-\frac{1}{18}\nabla_{\mu}S_{\nu}\nabla^{\mu}S^{\nu}-\frac{1}{36}\nabla_{\mu}S^{\mu}\nabla_{\nu}S^{\nu}\nonumber\\
    &+\frac{2}{9}\varepsilon^{\lambda\rho\mu\nu}\nabla_{\lambda}T_{\rho}\nabla_{\mu}S_{\nu}-\frac{16}{27}T^{\mu}T^{\nu}\nabla_{\mu}T_{\nu}+\frac{16}{27}T_{\mu}T^{\mu}\nabla_{\nu}T^{\nu}+\frac{1}{27}S^\mu S^\nu \nabla_\mu T_\nu-\frac{1}{27}S_{\mu}S^{\mu}\nabla_{\nu}T^{\nu}\nonumber\\
    &+\frac{1}{27}T^{\mu}S^{\nu}\nabla_{\mu}S_{\nu}+\frac{1}{27}T^\mu S^\nu\nabla_\nu S_\mu-\frac{2}{27}T_{\mu}S^{\mu}\nabla_{\nu}S^{\nu}+\frac{4}{27}T_{\mu}T^{\mu}T_{\nu}T^{\nu}+\frac{1}{1728}S_{\mu}S^{\mu}S_{\nu}S^{\nu} \nonumber\\
    &-\frac{1}{54} T_{\mu}T^{\mu}S_{\nu}S^{\nu}-\frac{1}{27}T_{\mu}S^{\mu}T_{\nu}S^{\nu}-\frac{8}{3}R_{\mu\nu}\nabla^{\mu}T^{\nu}+\frac{8}{9}G_{\mu\nu}T^{\mu}T^{\nu}-\frac{1}{18}G_{\mu\nu}S^{\mu}S^{\nu}\,,\label{R1}\\
    \tilde{R}_{\lambda\rho\mu\nu}\tilde{R}^{\mu\nu\lambda\rho}=&\;R_{\lambda\rho\mu\nu}R^{\lambda\rho\mu\nu}+\frac{8}{9}\nabla_{\mu}T_{\nu}\nabla^{\nu}T^{\mu}+\frac{4}{9}\nabla_{\mu}T^{\mu} \nabla_{\nu}T^{\nu}+\frac{1}{18}\nabla_{\mu}S_{\nu}\nabla^{\mu}S^{\nu}-\frac{1}{18}\nabla_{\mu}S^{\mu}\nabla_{\nu}S^{\nu}\nonumber\\
    &-\frac{2}{9}\varepsilon^{\lambda\rho\mu\nu}\nabla_{\lambda}T_{\rho}\nabla_{\mu}S_{\nu}-\frac{16}{27}T^{\mu}T^{\nu}\nabla_{\mu}T_{\nu}+\frac{16}{27}T_{\mu}T^{\mu}\nabla_{\nu}T^{\nu}+\frac{1}{27}S^{\mu}S^{\nu}\nabla_{\mu}T_{\nu}-\frac{1}{27}S_\mu S^{\mu}\nabla_{\nu}T^{\nu}\nonumber\\
    &-\frac{1}{27}T^\mu S^\nu\nabla_\mu S_\nu-\frac{1}{27}T^\mu S^\nu\nabla_\nu S_\mu -\frac{1}{27}T_\mu S^\mu\nabla_\nu S^\nu +\frac{4}{27}T_{\mu}T^{\mu}T_{\nu}T^{\nu}+\frac{1}{1728}S_{\mu}S^{\mu}S_{\nu}S^{\nu}\nonumber\\
    &+\frac{1}{162}T_{\mu}T^{\mu}S_{\nu}S^{\nu}-\frac{2}{81}T_{\mu}S^{\mu}T_{\nu}S^{\nu}-\frac{8}{3}R_{\mu\nu}\nabla^{\mu}T^{\nu}+\frac{8}{9}G_{\mu\nu}T^{\mu}T^{\nu}-\frac{1}{18}G_{\mu\nu}S^{\mu}S^{\nu}\,,\label{R2}\\
    \tilde{R}_{\lambda\rho\mu\nu}\tilde{R}^{\lambda\mu\rho\nu}=&\;\frac{1}{2}R_{\lambda\rho\mu\nu}R^{\lambda\rho\mu\nu}+\frac{2}{9}\nabla_{\mu}T_{\nu}\nabla^{\mu}T^{\nu}+\frac{2}{9}\nabla_{\mu}T_{\nu}\nabla^{\nu}T^{\mu}+\frac{2}{9}\nabla_{\mu}T^{\mu} \nabla_{\nu}T^{\nu} +\frac{{1}}{24}\nabla_{\mu}S^{\mu}\nabla_{\nu}S^{\nu}\nonumber\\
    &-\frac{8}{27}T^{\mu}T^{\nu}\nabla_{\mu}T_{\nu}+\frac{8}{27}T_{\mu}T^{\mu}\nabla_{\nu}T^{\nu}+\frac{1}{54}S^{\mu}S^{\nu}\nabla_{\mu}T_{\nu}-\frac{1}{54}S_{\mu}S^{\mu}\nabla_{\nu}T^{\nu}+\frac{3}{54}T_{\mu}S^{\mu}\nabla_{\nu}S^{\nu}\nonumber\\
    &+\frac{2}{27}T_{\mu}T^{\mu}T_{\nu}T^{\nu}+\frac{1}{3456}S_{\mu}S^{\mu}S_{\nu}S^{\nu}-\frac{1}{324}T_{\mu}T^{\mu}S_{\nu}S^{\nu}+\frac{1}{81} T_{\mu}S^{\mu}T_{\nu}S^{\nu}-\frac{4}{3}R_{\mu\nu}\nabla^{\mu}T^{\nu}\nonumber\\
    &+\frac{4}{9}G_{\mu\nu}T^{\mu}T^{\nu}-\frac{1}{36}G_{\mu\nu}S^{\mu}S^{\nu}\,,\label{R3}\\
    \tilde{R}_{\mu\nu}\tilde{R}^{\mu\nu}=&\;R_{\mu\nu}R^{\mu\nu}+\frac{4}{9}\nabla_{\mu}T_{\nu}\nabla^{\mu}T^{\nu}+\frac{8}{9}\nabla_{\mu}T^{\mu} \nabla_{\nu}T^{\nu}-\frac{1}{72}\nabla_{\mu}S_{\nu}\nabla^{\mu}S^{\nu}+\frac{1}{72}\nabla_{\mu}S_{\nu}\nabla^{\nu}S^{\mu}+\frac{1}{9}\varepsilon^{\lambda\rho\mu\nu}\nabla_{\lambda}T_{\rho}\nabla_{\mu}S_{\nu}\nonumber\\
    &-\frac{8}{27}T^{\mu}T^{\nu}\nabla_{\mu}T_{\nu}+\frac{20}{27}T_{\mu}T^{\mu}\nabla_{\nu}T^{\nu}+\frac{1}{54}S^{\mu}S^{\nu}\nabla_{\mu}T_{\nu}-\frac{5}{108}S_{\mu}S^{\mu}\nabla_{\nu}T^{\nu}\nonumber\\
    &+\frac{4}{27}T_{\mu}T^{\mu}T_{\nu}T^{\nu}+\frac{1}{1728}S_{\mu}S^{\mu}S_{\nu}S^{\nu}- \frac{1}{81}T_{\mu}T^{\mu}S_{\nu}S^{\nu}-\frac{1}{162}T_{\mu}S^{\mu}T_{\nu}S^{\nu}\nonumber\\
    &-\frac{2}{3}\left(2R_{\mu\nu}\nabla^{\mu}T^{\nu}+R\,\nabla_{\mu}T^{\mu}\right)+\frac{4}{9}\left(R_{\mu\nu}T^{\mu}T^{\nu}-R\,T_{\mu}T^{\mu}\right)-\frac{1}{36}\left(R_{\mu\nu}S^{\mu}S^{\nu}-R\,S_{\mu}S^{\mu}\right)\,,\label{R4}\\
    \tilde{R}_{\mu\nu}\tilde{R}^{\nu\mu}=&\;R_{\mu\nu}R^{\mu\nu}+\frac{4}{9}\nabla_{\mu}T_{\nu} \nabla^{\nu}T^{\mu}+\frac{8}{9}\nabla_{\mu}T^{\mu} \nabla_{\nu}T^{\nu}+\frac{1}{72}\nabla_{\mu}S_{\nu}\nabla^{\mu}S^{\nu}-\frac{1}{72}\nabla_{\mu}S_{\nu}\nabla^{\nu}S^{\mu}-\frac{1}{9}\varepsilon^{\lambda\rho\mu\nu}\nabla_{\lambda}T_{\rho}\nabla_{\mu}S_{\nu}\nonumber\\
    &-\frac{8}{27}T^{\mu}T^{\nu}\nabla_{\mu}T_{\nu}+\frac{20}{27}T_{\mu}T^{\mu}\nabla_{\nu}T^{\nu}+\frac{1}{54}S^{\mu}S^{\nu}\nabla_{\mu}T_{\nu}-\frac{5}{108}S_{\mu}S^{\mu}\nabla_{\nu}T^{\nu}\nonumber\\
    &+\frac{4}{27}T_{\mu}T^{\mu}T_{\nu}T^{\nu}+\frac{1}{1728}S_{\mu}S^{\mu}S_{\nu}S^{\nu}-\frac{1}{81}T_{\mu}T^{\mu}S_{\nu}S^{\nu}-\frac{1}{162}T_{\mu}S^{\mu}T_{\nu}S^{\nu}\nonumber\\
    &-\frac{2}{3}\left(2R_{\mu\nu}\nabla^{\mu}T^{\nu}+R\,\nabla_{\mu}T^{\mu}\right)+\frac{4}{9}\left(R_{\mu\nu}T^{\mu}T^{\nu}-R\,T_{\mu}T^{\mu}\right)-\frac{1}{36}\left(R_{\mu\nu}S^{\mu}S^{\nu}-R\,S_{\mu}S^{\mu}\right)\,,\label{R5}\\
    \tilde{R}^{2}=&\;R^{2}+4\nabla_{\mu}T^{\mu}\nabla_{\nu}T^{\nu}+\frac{8}{3}T_{\mu}T^{\mu}\nabla_{\nu}T^{\nu}-\frac{1}{6}S_{\mu}S^{\mu}\nabla_{\nu}T^{\nu}+\frac{4}{9}T_{\mu}T^{\mu}T_{\nu}T^{\nu}\nonumber\\
    &+\frac{1}{576}S_{\mu}S^{\mu}S_{\nu}S^{\nu}-\frac{1}{18}T_{\mu}T^{\mu}S_{\nu}S^{\nu}-4R\nabla_{\mu}T^{\mu}-\frac{4}{3}RT_{\mu}T^{\mu}+\frac{1}{12}RS_{\mu}S^{\mu}\,,\label{R6}
\end{align}
whereas the contribution of these modes to the three independent quadratic torsion scalars is:
\begin{align}
    T_{\lambda\mu\nu}T^{\lambda\mu\nu}&=\frac{2}{3}T_{\mu}T^{\mu}-\frac{1}{6}S_{\mu}S^{\mu}\,,\\
    T_{\lambda\mu\nu}T^{\mu\lambda\nu}&=\frac{1}{3}T_{\mu}T^{\mu}+\frac{1}{6}S_{\mu}S^{\mu}\,,\\
    T^{\lambda}{}_{\lambda\nu}T^{\mu}{}_{\mu}{}^{\nu}&=T_{\mu}T^{\mu}\,.
\end{align}

\section{First branch of the cubic Lagrangian in terms of the torsion tensor}\label{sec:AppCubic}

The Lagrangian~\eqref{cubicLagIrr} can be rewritten in terms of the torsion tensor as $\mathcal{\bar{L}}_{\rm curv-tors}^{(3)}=\mathcal{\bar{L}}^{(3)}_{1}+\mathcal{\bar{L}}^{(3)}_{2}+\mathcal{\bar{L}}^{(3)}_{3}$:
\begin{eqnarray}
\mathcal{\bar{L}}^{(3)}_{1}&=&\bar{h}_{1}^{} \tilde{R}_{\rho \tau \gamma \mu } T_{\alpha }{}^{\gamma \mu \
} T^{\alpha \rho \tau } + \bar{h}_{2}^{} \tilde{R}_{\rho \gamma \tau \mu } 
T_{\alpha }{}^{\gamma \mu } T^{\alpha \rho \tau } + \bar{h}_{3}^{} \tilde{R}_{\alpha \tau \gamma \mu } T^{\alpha \rho \tau } T_{\rho }{}^{
\gamma \mu } + \bar{h}_{4}^{} \tilde{R}_{\alpha \gamma \tau \mu } T^{\alpha 
\rho \tau } T_{\rho }{}^{\gamma \mu } + \bar{h}_{5}^{} \tilde{R}_{\tau \gamma \alpha \mu } T^{\alpha \rho \tau } T_{\rho }{}^{\gamma \mu }\nonumber\\
&&+ \bar{h}_{6}^{} \tilde{R}_{\gamma \mu \alpha \tau } T^{\alpha \rho \tau } T_{\rho }{}^{\gamma \mu }+ \bar{h}_{7}^{} \tilde{R}_{\rho \tau \gamma \mu } T^{\tau \gamma \mu } T^{\rho } + \bar{h}_{8}^{} \tilde{R}_{\rho \gamma \tau \mu } T^{\tau \gamma \mu } T^{\rho } + \bar{h}_{9}^{} \tilde{R}_{\tau \gamma \rho \mu } T^{\tau \gamma \mu } T^{\rho } + \bar{h}_{10}^{} \tilde{R}_{\gamma \mu \rho \tau } T^{\tau \gamma \mu } T^{\rho }\nonumber\\
&&+ \bar{h}_{11}^{} \tilde{R}_{\alpha \tau \gamma \mu } T^{\alpha \rho \tau } T^{\gamma }{}_{\rho }{}^{\mu } + \bar{h}_{12}^{} \tilde{R}_{\alpha \gamma \tau \mu } T^{\alpha \rho \tau } T^{\gamma }{}_{\rho }{}^{\mu } + \bar{h}_{13}^{} \tilde{R}_{\alpha \mu \tau \gamma } T^{\alpha \rho \tau } T^{\gamma }{}_{\rho }{}^{\mu } + \bar{h}_{14}^{} \tilde{R}_{\tau \mu \alpha \gamma } T^{\alpha \rho \tau } T^{\gamma }{}_{\rho }{}^{\mu } \,,\\
\mathcal{\bar{L}}^{(3)}_{2}&=&\bar{h}_{15}^{} \tilde{R}^{\alpha \rho } T_{\alpha }{}^{\tau \gamma } T_{\rho \tau \gamma } + \bar{h}_{16}^{} \tilde{R}^{\alpha \rho } T_{\rho }{}^{\tau \gamma } T_{\tau \alpha \gamma } + \bar{h}_{17}^{} \tilde{R}^{\alpha \rho } T_{\alpha }{}^{\tau \gamma } T_{\tau \rho \gamma } + \bar{h}_{18}^{} \tilde{R}^{\alpha \rho } T_{\tau \rho \gamma } T^{\tau }{}_{\alpha }{}^{\gamma } + \bar{h}_{19}^{} \tilde{R}^{\alpha \rho } T^{\tau }{}_{\alpha }{}^{\gamma } T_{\gamma \rho \tau } \nonumber\\
&&+ \bar{h}_{20}^{} \tilde{R}^{\alpha \rho } T_{\alpha } T_{\rho } + \bar{h}_{21}^{} \tilde{R}^{\alpha \rho } T_{\alpha \rho }{}^{\tau } T_{\tau } + \bar{h}_{22}^{} \tilde{R}^{\alpha \rho } T_{\rho \alpha }{}^{\tau } T_{\tau } + \bar{h}_{23}^{} \tilde{R}^{\alpha \rho } T^{\tau }{}_{\alpha \rho } T_{\tau }\,,\\
\mathcal{\bar{L}}^{(3)}_{3}&=&\bar{h}_{24}^{} \tilde{R} T_{\alpha \rho \tau } T^{\alpha \rho \tau } + 
\bar{h}_{25}^{} \tilde{R} T^{\alpha \rho \tau } T_{\rho \alpha \tau } + 
\bar{h}_{26}^{} \tilde{R} T_{\rho } T^{\rho }\,,
\end{eqnarray}
where the respective Lagrangian coefficients are related as
    \begin{eqnarray}
        \bar{h}_{1}^{} &=& \frac{1}{3}\left(h_{5}{}+6 h_{20}{}-h_{7}{}\right)\,,\quad \bar{h}_{2}^{} = \frac{1}{3}\left(h_{6}{}+6 h_{24}{}-h_{8}{}-12 h_{20}{}\right)\,,\quad \bar{h}_{3}^{} =\frac{1}{3}\left(2h_{5}{}-6h_{21}{}\right)\,,\\
        \bar{h}_{4}^{} &=& \frac{1}{3}\left(2 h_{6}{}+2h_{9}{}+12h_{20}{} + 12 h_{22}{}-h_{8}{}- 6 h_{23}{} - 6 h_{24}{}\right)\,,\quad \bar{h}_{5}^{} = 2 h_{24}{}-4 h_{20}{} - 4 h_{21}{} -  \frac{2}{3} h_{6}{} -  \frac{1}{3} h_{8}{}\,,\\
        \bar{h}_{6}^{} &=& \frac{2}{3}\left(h_{5}{}-6h_{20}{} - 3h_{22}{}\right)\,,\quad \bar{h}_{7}^{} =\frac{2}{9} \left(h_{7}{} +  h_{8}{}+ 9 
h_{21}{} + 3 h_{24}{}- 2 h_{5}{} - h_{6}{}-18 h_{13}{} - 3 h_{17}{} - 12 h_{20}{}\right)\,,\\
\bar{h}_{8}^{} &=& \frac{2}{9} (36 h_{13}{} - 3 h_{17}{} + 24 h_{20}{} - 18 
h_{21}{} - 6 h_{24}{} - 2 h_{5}{} -  h_{6}{} + h_{7}{} + h_{8}{})\,,\\
\bar{h}_{9}^{} &=& \frac{1}{9}\left(4 h_{5}{} + 2 
h_{6}{}-72 h_{13}{} - 36 h_{14}{} - 3 h_{16}{} - 24 
h_{20}{} + 36 h_{22}{} - 12 h_{23}{} + 12 h_{24}{} - 2 h_{7}{} - 2 h_{9}{}\right)\,,\\
\bar{h}_{10}^{} &=& \frac{1}{9}\left(2 h_{7}{} + 2 h_{9}{}-36 h_{13}{} - 18 h_{14}{} + 3 h_{16}{} - 
12 h_{20}{} + 18 h_{22}{} - 6 h_{23}{} + 6 h_{24}{} - 4 h_{5}{} - 2 
h_{6}{}\right)\,,\\
\bar{h}_{11}^{} &=& 4 h_{21}{} + 4 h_{22}{} -  \frac{2}{3} h_{7}{}\,,\quad \bar{h}_{12}^{} = \frac{1}{3}\left(h_{9}{}+ 6 h_{23}{}-h_{8}{}-12 h_{22}{}\right)\,,\\
 \bar{h}_{13}^{} &=& \frac{1}{3} (12 h_{21}{} + 12 h_{22}{} - 6 h_{23}{} + 
h_{8}{} + 2 h_{9}{})\,,\quad \bar{h}_{14}^{} = -\,4h_{21}{}\,,\\
\bar{h}_{15}^{} &=& \frac{1}{9} (18 h_{3}{} + 2 h_{5}{} + h_{6}{} + 2 h_{7}{} 
+ h_{8}{} + h_{9}{}-2 h_{10}{} + h_{11}{} - 6 h_{23}{} - 6 
h_{24}{} + 6 h_{25}{})\,,\\
\bar{h}_{16}^{} &=& \frac{1}{9} \bigl[h_{11}{}-2 h_{10}{}+ 2 (6 h_{23}{}
+ 6 h_{24}{}- 18 h_{3}{} - 2 h_{5}{} -  
h_{6}{} - 2 h_{7}{} -  h_{8}{} -  h_{9}{}- 6 h_{25}{} - 18 h_{26}{})\bigr]\,,\\
\bar{h}_{17}^{} &=& \frac{1}{9}\left(h_{11}{}-2 h_{10}{} + 12 h_{23}{} + 12 
h_{24}{} + 6 h_{25}{} + 36 h_{26}{} - 36 h_{3}{} - 4 h_{5}{} - 2 
h_{6}{} - 4 h_{7}{} - 2 h_{8}{} - 2 h_{9}{}\right)\,,\\
\bar{h}_{18}^{} &=& \frac{1}{9} \bigl[5 h_{11}{}- h_{10}{} + 2\bigl(18 h_{3}{} + 2 h_{5}{} + h_{6}{} 
+ 2 h_{7}{} + h_{8}{} + h_{9}{}-6 
h_{23}{} - 6 h_{24}{} - 3 h_{25}{}\bigr)\bigr]\,,\\
\bar{h}_{19}^{} &=& \frac{1}{9} (h_{10}{} + 4 h_{11}{} + 12 h_{23}{} + 12 
h_{24}{} + 6 h_{25}{} - 36 h_{3}{} - 4 h_{5}{} - 2 h_{6}{} - 4 
h_{7}{} - 2 h_{8}{} - 2 h_{9}{})\,,\\
\bar{h}_{20}^{} &=& \frac{1}{9} (9 h_{1}{} + h_{10}{} - 4 h_{11}{} - 3 
h_{16}{} + 6 h_{17}{} - 3 h_{19}{} + 4 h_{5}{} + 2 h_{6}{} - 2 
h_{7}{} -  h_{8}{} - 2 h_{9}{})\,,\\
\bar{h}_{21}^{} &=& \frac{1}{9}\left(3 h_{11}{} + 18 h_{15}{}-3 h_{10}{} - 3 
h_{18}{} + 6 h_{19}{} + 12 h_{21}{} - 12 h_{22}{} + 6 h_{23}{} + 6 
h_{25}{} + 24 h_{26}{} -  h_{8}{} + 4 h_{9}{}\right)\,,\\
\bar{h}_{22}^{} &=& \frac{1}{9}\left(3 h_{11}{}-3 h_{10}{} - 18 h_{15}{} + 3 
h_{18}{} + 3 h_{19}{} - 12 h_{21}{} + 12 h_{22}{} - 6 h_{23}{} - 6 
h_{25}{} - 24 h_{26}{} + h_{8}{} + 2 h_{9}{}\right)\,,\\
\bar{h}_{23}^{} &=& \frac{1}{9} (18 h_{15}{} + 6 h_{18}{} - 3 h_{19}{} + 12 
h_{21}{} - 12 h_{22}{} + 6 h_{23}{} + 6 h_{25}{} + 24 h_{26}{} + 2 
h_{8}{} - 2 h_{9}{})\,,\\
\bar{h}_{24}^{} &=& \frac{1}{18} (12 h_{12}{} + 6 h_{23}{} + 6 h_{24}{} - 36
h_{3}{} - 36 h_{4}{} - 2 h_{5}{} -  h_{6}{} - 2 h_{7}{} -  h_{8}{} -  
h_{9}{})\,,\\
\bar{h}_{25}^{} &=& \frac{1}{9}\left(36 
h_{3}{} + 36 h_{4}{} + 2 h_{5}{} + h_{6}{} + 2 h_{7}{} + h_{8}{} + 
h_{9}{}+6 h_{12}{}- 6 h_{23}{} - 6 h_{24}{}\right)\,,\\
\bar{h}_{26}^{} &=& \frac{1}{9}\left(9 h_{2}{} + h_{9}{}+ h_{11}{}+ 3 
h_{19}{}- h_{10}{} - 6 h_{12}{}\right)\,.
    \end{eqnarray}

\newpage

\bibliographystyle{utphys}
\bibliography{references}

\end{document}